\documentclass[aps,showpacs,nofootinbib,superscriptaddress]{revtex4}

\usepackage{graphicx}
\usepackage{dcolumn}

\def\slashchar#1{\setbox0=\hbox{$#1$}
   \dimen0=\wd0 \setbox1=\hbox{/} \dimen1=\wd1
   \ifdim\dimen0>\dimen1 \rlap{\hbox to \dimen0{\hfil/\hfil}} #1
   \else  \rlap{\hbox to \dimen1{\hfil$#1$\hfil}} / \fi}
\def\tstrut{\vrule height2.5ex depth0pt width0pt} 
\def\esp{\kern 9mm} 
\begin{document}
\title{Inclusive Nucleon Emission Induced by Quasi--Elastic
  Neutrino--Nucleus Interactions.}
\author{J. Nieves}
\affiliation{Departamento de F\'\i sica Moderna, \\Universidad de
Granada, E-18071 Granada, Spain} 
\author{M. Valverde}
\affiliation{Departamento de F\'\i sica Moderna, \\Universidad de
Granada, E-18071 Granada, Spain} 
\author{M.J. Vicente Vacas}
\affiliation{Departamento de F\'\i sica Te\'orica and IFIC,
Centro Mixto Universidad de Valencia-CSIC\\ Institutos de Investigaci\'on
de Paterna, Aptdo. 22085, E-46017 Valencia, Spain}
\today
\begin{abstract}
\rule{0ex}{3ex} We study the quasi--elastic contribution to the
inclusive $(\nu_l,\nu_l N)$, $(\nu_l,l^- N)$, $({\bar \nu}_l,{\bar
\nu}_l N)$ and $({\bar \nu}_l,l^+ N)$ reactions in nuclei using a
Monte Carlo simulation method to account for the rescattering of the
outgoing nucleon. As input, we take the reaction probability from the
microscopical many body framework developed in Phys.\ Rev.\ {\bf C70}
(2004) 055503 for charged-current induced reactions, while for neutral
currents we use results from a natural extension of the model
described in that reference.  The nucleon emission process studied
here is a clear signal for neutral--current neutrino driven reactions, that 
can be used in the analysis of future neutrino experiments.

\end{abstract}


\maketitle



\section{Introduction}

Neutrino physics is at the forefront of current theoretical and
experimental research in astro, nuclear, and particle physics. Indeed,
neutrino interactions offer unique opportunities for exploring
fundamental questions in these domains of the physics. One of these
questions is the neutrino-oscillation phenomenon, for which there have
been conclusive positive signals in the last
years~\cite{osc}. Neutrino-oscillation experiments are presently
evolving from the discovery to the precision phase. This new
generation of experiments faces a major difficulty: the elusive
nature of the neutrinos. The presence of neutrinos, being chargeless
particles, can only be inferred by detecting the secondary particles
they create when colliding and interacting with matter. Nuclei are
often used as neutrino detectors, thus a trustable interpretation of
neutrino data heavily relies on  detailed and quantitative knowledge
of the features of the neutrino-nucleus interaction~\cite{RCNN}. For
instance, in the case of neutrino processes driven by the electroweak
Neutral Current (NC), the energy spectrum and angular distribution of
the ejected nucleons are the unique observables. There is a general
consensus among the theorists that a simple Fermi gas model, widely
used in the analysis of neutrino oscillation experiments, fails to
provide a satisfactory description of the measured cross sections, and
inclusion of further nuclear effects is needed~\cite{Benhar:2005dj}.

Simultaneously, in recent years there have also been some initiatives
aiming at understanding the quark and gluon substructure of the
nucleon.  The flavor dependence of the nucleon's spin is a significant
fundamental question that is not yet fully understood.
Experiments~\cite{emc, ah87} measuring the spin content of the nucleon
have reported conflicting results on the amount of nucleon spin
carried by strange quarks~\cite{ej74}.  Recently, the FINeSSE
collaboration at Fermilab has suggested~\cite{finese,Pate:2005ft} that
Quasi-Elastic (QE) neutrino--nucleus scattering, observed using a
novel detection technique, provides a theoretically clean measure of
this quantity. In this context, it is also necessary to control
nuclear effects.

 At intermediate energies, above the nuclear giant resonance and below
 the $\Delta(1232)$ regions\footnote{There exists an abundant
 literature studying these two regions. See for instance a recent
 paper~\cite{BoCo05} on the excitation of nuclear giant resonances in
 neutrino scattering off nuclei or the older works of
 Refs.~\cite{Ko97}--\cite{Ko03} also studying QE neutrino--nucleus
 scattering at low energies.  On the $\Delta-$excitation in neutrino
 reactions, there are works, among others, by Alvarez-Ruso and
 collaborators~\cite{aruso}, S.K. Singh and
 collaborators~\cite{Singh:1998ha} and Lalakulich y
 Paschos~\cite{Lalakulich:2005cs}.}, neutrino--nucleus interactions
 have been studied within several approaches. Several different Fermi
 gas, Random Phase Approximation (RPA), shell model and superscaling
 based calculations have been developed during the last 15
 years~\cite{Garvey:1992qp}--\cite{superscal}.  Some of these
 approaches have been also employed to compute neutrino or
 antineutrino induced single--nucleon emission cross sections. Most of
 the calculations use the plane wave and distorted wave impulse
 approximations (PWIA and DWIA, respectively), including or not
 relativistic effects. The PWIA calculations neglect all types of
 interactions between the ejected nucleon and the residual nuclear
 system, and therefore such a framework constitutes a poor
 approximation to evaluate nucleon emission cross sections. However,
 the PWIA has been often used to compute the ratio of proton $(\nu,p)$
 to neutron $(\nu,n)$ yields, which at low neutrino energies and for
 light nuclei might be rather insensitive to rescattering effects.

 Within the DWIA the ejected nucleon is described as a solution to the
 Dirac or Schr\"odinger equation with an optical potential obtained by
 fitting elastic proton--nucleus scattering data. The imaginary part
 accounts for the absorption into unobserved channels\footnote{For
 nucleon energies above 1 GeV, the Glauber
 model~\cite{Glauber:1970jm}, which is a multiple--scattering
 extension of the eikonal approximation, has also been used (see for
 instance the recent work of Ref.~\cite{Martinez:2005xe}). In this
 approach, a relativistic plane wave is modulated by a factor which
 accounts for the absorption into unobserved channels. }. This scheme,
 first developed in $(e,e^\prime p)$ studies where the final nucleus
 is left in the ground or in a particular excited state, is incorrect
 to study nucleon emission processes where the state of the final
 nucleus is totally unobserved, and thus all final nuclear
 configurations, either in the discrete or on the continuum,
 contribute.  The distortion of the nucleon wave function by a complex
 optical potential removes all events where the nucleons collide with
 other nucleons. Thus, in DWIA calculations, the nucleons that
 interact are lost when in the physical process they simply come off
 the nucleus with a different energy, angle, and maybe charge, and
 they should definitely be taken into account. A clear example which
 illustrates the deficiencies of the DWIA models is the neutron
 emission process: $(\nu_l, l^- n)$. Within the impulse approximation
 neutrinos only interact via Charged Current (CC) interactions with
 neutrons and would emit protons, and therefore the DWIA will predict zero
 cross sections for CC one neutron knock-out reactions. However, the
 primary protons interact strongly with the medium and collide with
 other nucleons which are also ejected. As a consequence there is a
 reduction of the flux of high energy protons but a large number of
 secondary nucleons, many of them neutrons, of lower energies appear.

 The distortion by a real potential does not eliminate the events
 where there are nucleon collisions. But, it does not account either
 for the changes of energy, direction and charge of the nucleons
 induced by these collisions. Besides, it cannot account either for more
 than one nucleon knock-out events.

In this work, we study the QE contribution to the inclusive
$(\nu_l,\nu_l N)$, $(\nu_l,l^- N)$, $({\bar \nu}_l,{\bar \nu}_l N)$
and $({\bar \nu}_l,l^+ N)$ reactions in nuclei. We use a Monte Carlo
(MC) simulation method to account for the rescattering of the outgoing
nucleon.  A reliable description of the gauge bosons ($W^{\pm}$ and
$Z^0$ ) absorption in the nucleus is the first essential ingredient.

For CC driven processes, we use the many body framework developed in
Ref.~\cite{NAV05}. Starting from a Local Fermi Gas (LFG) picture of
the nucleus, which automatically accounts for Pauli blocking, several
nuclear effects are taken into account in the scheme of
Ref.~\cite{NAV05}: {\it i)} a correct energy balance, using the
experimental $Q-$values, is enforced, {\it ii)} Coulomb distortion of
the charged leptons is implemented by using the so called ``modified
effective momentum approximation''~\cite{En98}, {\it iii)} medium
polarization (RPA), including $\Delta-$hole degrees of freedom and
explicit pion and rho exchanges in the vector--isovector channel of
the effective nucleon--nucleon force, and Short Range Correlation
(SRC) effects are computed, and finally {\it iv)} the nucleon
propagators are dressed in the nuclear medium, which amounts to work
with a LFG of interacting nucleons and it also accounts for reaction
mechanisms where the gauge boson, $W^+$ or $W^-$, is absorbed by two
nucleons (the real part of the nucleon selfenergy modifies the free
nucleon dispersion relation, while the imaginary part takes into
account two nucleon absorption reaction channels).  The model has no
free parameters. The $W^{\pm}N$ couplings and form factors are fixed
in the vacuum, while the main features concerning the nuclear
corrections, expansion parameter and all sorts of constants, are
completely fixed from previous hadron-nucleus studies (pionic atoms,
elastic and inelastic pion-nucleus reactions, $\Lambda-$ hypernuclei,
etc.)~\cite{OTW82,pion,pion1}. Thus, the model is a natural extension
of previous studies~\cite{GNO97, CO92,OTW82,pion,pion1} on electron,
photon and pion dynamics in nuclei, and should be able to describe
inclusive CC QE neutrino and antineutrino nuclear reactions at
intermediate energies of interest for future neutrino oscillation
experiments. Even though the scarce existing CC data involve very low
nuclear excitation energies, for which specific details of the nuclear
structure might play an important role, the model of Ref.~\cite{NAV05}
provides one of the best existing combined description of the
inclusive muon capture in $^{12}$C and of the measurements of the
$^{12}$C $(\nu_\mu,\mu^-)X$ and $^{12}$C $(\nu_e,e^-)X$ reactions near
threshold.  Inclusive muon capture from other nuclei is also
successfully described by the model.  Besides, above, let us say 80 or
100 MeV of energy transferred to the nucleus, this many body framework
leads also to excellent results for the $(e,e^\prime)$ inclusive
reaction in nuclei, not only in the QE region, but also when it is
extended to the study of the $\Delta-$peak and the dip region
(situated between the QE and the $\Delta$
peaks)~\cite{GNO97}\footnote{Data in $^{12}$C, $^{40}$Ca and
$^{208}$Pb of differential cross sections for different electron
kinematics and split into longitudinal and transverse response
functions are successfully described. } and to the description of the
absorption of real photons by nuclei~\cite{CO92}.

In Sect.~\ref{sec:nc}, we extend the model of
Ref.~\protect\cite{NAV05} to NC driven processes, both for neutrino
and antineutrino induced nuclear reactions in the QE region. Thus we
 compute, for a fixed incoming neutrino or
antineutrino Laboratory (LAB) energy, the inclusive QE cross sections
$d^2\sigma /d\Omega^{\prime} dE^{\prime}$ ($\Omega^{\prime}$,
$E^{\prime}$ are the solid angle and energy of the outgoing lepton)
for $(\nu_l,\nu_l)$, $(\nu_l,l^-)$, $({\bar \nu}_l,{\bar \nu}_l)$ and
$({\bar \nu}_l,l^+ )$ processes. This cross section gives us the
reaction probability and it is the first required ingredient to start
with our cascade model to describe the collisions suffered by the
nucleons through their way out of the nucleus\footnote{Besides, we
 also compute differential cross sections with respect to $d^3r$. Thus, we
also know the point of the nucleus where the gauge boson was absorbed,
and we can start from there our MC propagation of the ejected
nucleon.}. Details on the MC simulation are given in
Sect.~\ref{sec:MC}. The MC method used here was designed  for
single and multiple nucleon and pion emission reactions induced by
pions~\cite{pion,VicenteVacas:1993bk}  and has been successfully employed to
describe inclusive $(\gamma, \pi)$, $(\gamma, N)$, $(\gamma, NN)$,...,
$(\gamma, N\pi)$,...~\cite{CO92bis,COV94}, $(e, e^\prime \pi)$, $(e,
e^\prime N)$, $(e, e^\prime NN)$,..., $(e, e^\prime
N\pi)$,...~\cite{GNO97bis} reactions in nuclei or the neutron and
proton spectra from the decay of $\Lambda$
hypernuclei~\cite{Ramos:1996ik}. Thus, we are using a quite robust and
well tested MC simulator.

In Sect.~\ref{sec:RS} we discuss our results and the main conclusions
of this work are outlined in Sect.~\ref{sec:concl}. We start
presenting (Subsect.~\ref{sec:RSNC}) results for the inclusive QE NC
cross sections for both neutrino and antineutrino beams in several
nuclei and in the next subsection we show results for inclusive
$(\nu_l,\nu_l N)$, $(\nu_l,l^- N)$, $({\bar \nu}_l,{\bar \nu}_l N)$
and $({\bar \nu}_l,l^+ N)$ reactions in nuclei at low energies,
obtained from our cascade model. Some preliminary results of this work
were already presented in Ref.~\cite{valverde}.  Finally, in the
appendix we give explicit expressions for the NC nucleon tensor, both
in the impulse approximation and when RPA corrections are taken into
account.

To end this introduction, we would like to devote a few words on the
applicability of the nuclear model used here. One might think that a LFG
description of the nucleus is poor, and that a proper finite nuclei
treatment is necessary. For inclusive processes and nuclear excitation
energies of around 100 MeV or higher, the findings of
Refs.~\cite{pion1}, \cite{GNO97} and \cite{CO92} clearly contradict
this conclusion. The reason is that in these circumstances one should
sum up over several nuclear configurations, both in the discrete and
in the continuum, and this inclusive sum is almost not sensitive to the
details of the nuclear wave function\footnote{The results of
Ref.~\cite{NAV05} for the inclusive muon capture in nuclei through the
whole periodic table, where the capture widths vary from about
4$\times 10^4$ s$^{-1}$ in $^{12}$C to 1300 $\times 10^4$ s$^{-1}$ in
$^{208}$Pb, and of the LSND measurements of the $^{12}$C
$(\nu_\mu,\mu^-)X$ and $^{12}$C $(\nu_e,e^-)X$ reactions near
threshold indicate that the predictions of our scheme, for totally
integrated inclusive observables, could even be extended to much
smaller, of the order of 10 or 20 MeV, nuclear excitation energies.
In this respect, the works of Refs.~\cite{mucap} and ~\cite{picap} for
inclusive muon capture and radiative pion capture in nuclei,
respectively, turn out to be quite enlightening. In those works,
continuum shell model results are compared to those obtained from a
LFG model for several nuclei from $^{12}$C to $^{208}$Pb. The
differential decay width shapes predicted for the two  set of models
are substantially different. Shell model distributions
present discrete contributions and in the continuum appear sharp
scattering resonances. Despite the fact that those distinctive 
features do not appear
in the LFG differential decay widths, the totally integrated widths
(inclusive observable) obtained from both descriptions of the
process do not differ in more than 5 or 10\%. The typical
nuclear excitation energies in muon and radiative pion capture in
nuclei are small, of the order of 20 MeV, and thus one expects that at
higher excitation energies, where one should sum up over a larger
number of nuclear final states, the LFG predictions for inclusive
observables would become even more reliable.}, in sharp contrast to
what happens in the case of exclusive processes where the final
nucleus is left in a determined nuclear level. On the other hand, the
LFG description of the nucleus allows for an accurate treatment of the
dynamics of the elementary processes (interaction of gauge bosons with
nucleons, nucleon resonances, and mesons, interaction between nucleons
or between mesons and nucleons, etc.)  which occur inside the nuclear
medium. Within a finite nuclei scenario, such a treatment becomes
hard to implement, and often the dynamics is simplified in order to
deal with more elaborated nuclear wave functions. This simplification
of the dynamics cannot lead to a good description of nuclear inclusive
electroweak processes at the intermediate energies of interest for
future neutrino experiments.

\section{Extension of the model of Ref.~\protect\cite{NAV05} to
  neutral--currents}\label{sec:nc}

\subsection{General formalism: hadronic tensor and many body expansion}

We will first focus  on the neutrino induced inclusive reaction
driven by the electroweak NC
\begin{equation}
\nu_l(k) + A_Z \to \nu_l (k^\prime)+ X
\label{eq:reac}
\end{equation}
and we will follow the same notation and convention as in
Ref.~\cite{NAV05}. The double differential cross section, with respect
to the outgoing neutrino kinematical variables, for the process of
Eq.~(\ref{eq:reac}) and for massless neutrinos, is given in the
LAB frame by
\begin{eqnarray}
\frac{d^2\sigma_{\nu \nu}}{d\Omega(\hat{k^\prime})d|\vec{k}^\prime|} &=&
\frac{ |\vec{k}^\prime|^2  M_iG^2}{4\pi^2} \left \{ 2W_1
\sin^2\frac{\theta^\prime}{2} + W_2
\cos^2\frac{\theta^\prime}{2} - W_3 \frac{|\vec{k}\,|+|\vec{k}^\prime|}{M_i}
\sin^2\frac{\theta^\prime}{2}\right\} \label{eq:cross}
\end{eqnarray}
with $\vec{k}$ and $\vec{k}^\prime~$ the LAB neutrino momenta,
$G=1.1664\times 10^{-11}$ MeV$^{-2}$, the Fermi constant,
$\theta^\prime$ the outgoing neutrino scattering angle and $M_i$ the
target nucleus mass. To obtain Eq.~(\ref{eq:cross}) we have neglected
the four-momentum carried out by the intermediate $Z-$boson with
respect to its mass. Finally, the three independent, Lorentz scalar
and real, structure functions, $W_i(q^2)$, enter into the definition
of  the hadronic
tensor, $W^{\mu\nu}$, which includes all sort of non-leptonic vertices
and corresponds to the neutral current electroweak transitions of the
target nucleus, $i$, to all possible final states.  It is thus given
by (in our convention, we take $\epsilon_{0123}= +1$ and the metric
$g^{\mu\nu}=(+,-,-,-)$)\footnote{Note that: (i) Eq.~(\ref{eq:wmunu})
holds with states normalized so that $\langle \vec{p} |
\vec{p}^{\,\prime} \rangle = (2\pi)^3 2p_0
\delta^3(\vec{p}-\vec{p}^{\,\prime})$, (ii) the sum over final states
$f$ includes an integration $ \int \frac{d^3p_j}{(2\pi)^3 2E_j}$, for
each particle $j$ making up the system $f$, as well as a sum over all
spins involved.}
\begin{eqnarray}
\frac{W^{\mu\sigma}}{2M_i} &=& \frac{1}{4M_i^2}\overline{\sum_f } (2\pi)^3
\delta^4(P^\prime_f-P-q) \langle f | j^\mu_{\rm nc}(0) | i \rangle
 \langle f | j^\sigma_{\rm nc}(0) | i \rangle^* \nonumber \\
&&\nonumber \\
&=& - g^{\mu\nu}W_1 + \frac{P^\mu
  P^\nu}{M_i^2} W_2 + {\rm i}
  \frac{\epsilon^{\mu\nu\gamma\delta}P_\gamma q_\delta}{2M_i^2}W_3 +  
\frac{q^\mu  q^\nu}{M_i^2} W_4 + \frac{P^\mu q^\nu + P^\nu q^\mu}
{2M_i^2} W_5+ {\rm i}\frac{P^\mu q^\nu - P^\nu q^\mu}
{2M_i^2} W_6
\label{eq:wmunu}
\end{eqnarray}
with $P^\mu$ the four-momentum of the initial nucleus ($M_i^2=P^2$), 
$P_f^\prime$  the total four momentum of
the hadronic state $f$ and $q=k-k^\prime$ the four momentum
transferred to the nucleus.  The bar over the sum denotes the
average over initial spins. By construction, the hadronic
tensor accomplishes
\begin{eqnarray}
W^{\mu\sigma}= W^{\mu\sigma}_s + {\rm i} W^{\mu\sigma}_a 
\end{eqnarray}
with $W^{\mu\sigma}_s$ ($W^{\mu\sigma}_a$) real symmetric
(antisymmetric) tensors, and finally for the NC we take
\begin{equation}
j^\mu_{\rm nc} = \overline{\Psi}_u\gamma^\mu(1-\frac83 \sin^2\theta_W-
\gamma_5)\Psi_u  - \overline{\Psi}_d\gamma^\mu(1+\frac43 \sin^2\theta_W-
\gamma_5)\Psi_d  - \overline{\Psi}_s\gamma^\mu(1+\frac43 \sin^2\theta_W-
\gamma_5)\Psi_s  
\end{equation}
with $\Psi_u$, $\Psi_d$ and $\Psi_s$ quark fields, and $\theta_W$ the
Weinberg angle ($\sin^2\theta_W= 0.231$).

Taking $\vec{q}$ in the $z$ direction, ie, $\vec{q}= |q| {\vec u}_z$,
and $P^\mu = (M_i, \vec{0})$, it is straightforward to find the six
structure functions in terms of the $W^{00}, W^{xx}=W^{yy}, W^{zz},
W^{xy}$ and $W^{0z}$ components of the hadronic tensor\footnote{ Thus,
  one readily finds
\begin{equation}
W_1=\frac{W^{xx}}{2M_i}, \quad W_2=\frac{1}{2M_i} \left (W^{00}+W^{xx}
+ \frac{(q^0)^2}{|\vec{q}\,|^2}(W^{zz}-W^{xx})-
2\frac{q^0}{|\vec{q}\,|}{\rm Re}~W^{0z}\right), \quad 
W_3=-{\rm i}\frac{W^{xy}}{|\vec{q}\,|} \label{eq:ws}
\end{equation}}.   The neutrino cross section, Eq.~(\ref{eq:cross}), does not
depend on $M_i$, as can be seen from the relations of
Eq.~(\ref{eq:ws}), and also note that the structure functions
$W_{4,5,6}$ do not contribute in the limit of massless neutrinos.

The cross section for  the antineutrino induced nuclear reaction
\begin{equation}
{\bar \nu}_l (k) +\, A_Z \to  {\bar \nu}_l (k^\prime) + X  \label{eq:anti}
\end{equation}
is easily obtained from Eq.~(\ref{eq:cross}), just by changing the
sign of the parity-violating term ($W_3$).

The hadronic tensor is determined by the $Z^0-$boson selfenergy,
$\Pi^{\mu\rho}_Z(q)$, in the nuclear medium. We follow here the
formalism of Ref.~\cite{NAV05}, and we evaluate the selfenergy,
$\Sigma_\nu^r(k;\rho)$, of a neutrino, with four-momentum $k$ and
helicity $r$, moving in infinite nuclear matter of constant density
$\rho$. We find,
\begin{eqnarray}
W^{\mu\sigma}_s &=& - \Theta(q^0) \left (\frac{4\cos\theta_W}{g} \right )^2 
\int \frac{d^3 r}{2\pi}~ {\rm Im}\left [ \Pi_Z^{\mu\sigma} 
+ \Pi_Z^{\sigma\mu} \right ] (q;\rho(r))\label{eq:wmunus}\\
W^{\mu\sigma}_a &=& - \Theta(q^0) \left (\frac{4\cos\theta_W}{g} \right )^2
 \int \frac{d^3 r}{2\pi}~{\rm Re}\left [ \Pi_Z^{\mu\sigma} 
- \Pi_Z^{\sigma\mu}\right] (q;\rho(r)), \label{eq:wmunua}
\end{eqnarray}
where we have used the Local Density Approximation (LDA) to obtain
results in a finite nucleus of density $\rho(r)$, and $g$ is the gauge
weak coupling constant, $g = e/\sin \theta_W$, related to the Fermi constant:
$G/\sqrt 2 = g^2/8M^2_W$, with $e$ the electron charge.

\begin{figure}[tbh]
\centerline{\includegraphics[height=12.0cm]{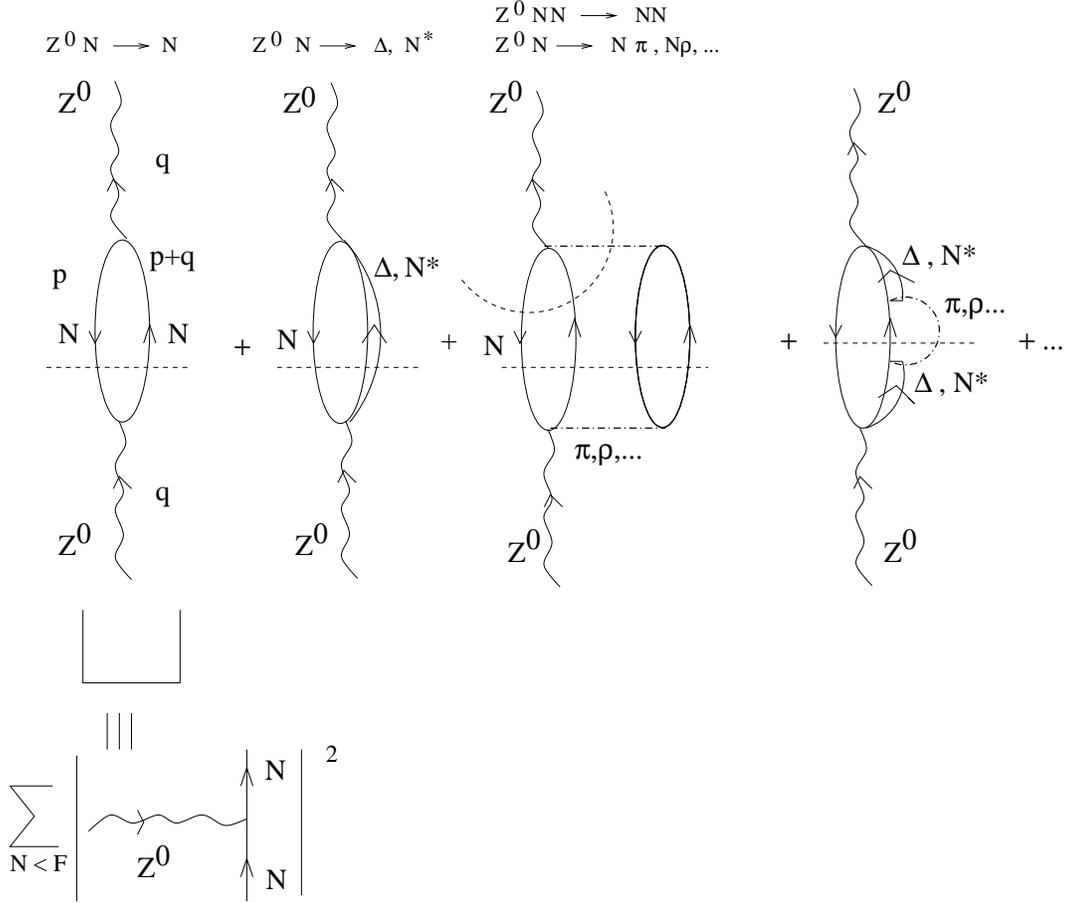}}
\caption{ Diagrammatic representation of some diagrams
  contributing to the $Z^0-$selfenergy. }\label{fig:fig2}
\end{figure}

 As we see, the basic object is the selfenergy of the Gauge Boson
($Z^{0}$) inside of the nuclear medium. As it is done in
Ref.~\cite{GNO97} for electro induced nuclear reactions, 
we plan to perform a many body expansion, where the
relevant gauge boson absorption modes would be systematically
incorporated: absorption by one nucleon, or a pair of nucleons or even
three nucleon mechanisms, real and virtual meson ($\pi$, $\rho$,
$\cdots$) production, excitation of $\Delta$ of higher resonance
degrees of freedom, etc. In addition, nuclear effects such as RPA or
SRC should also be taken
into account.  Some of the $Z^0-$absorption modes are depicted in
Fig.~\ref{fig:fig2}.

\subsection{QE contribution and Pauli blocking}

The virtual $Z^0$ can be absorbed by one nucleon leading to the QE
contribution of the nuclear response function. Such a contribution
corresponds to a 1p1h nuclear excitation (first of the diagrams
depicted in Fig.~\ref{fig:fig2}). To evaluate this selfenergy, 
the free nucleon propagator in the medium is required.
\begin{equation}
S(p\,; \rho) = (\slashchar{p}+M) G(p\,; \rho),
\qquad G(p\,; \rho) = \left ( \frac{1}{p^2-M^2+ {\rm
    i}\epsilon}  +  \frac{2\pi{\rm i}}{2E(\vec{p}\,)}
    \delta(p^0-E(\vec{p}\,)) \Theta(k_F-|\vec{p}~|) \right ) \label{eq:Gp}
\end{equation}
with the local Fermi momentum $k_F(r)= (3\pi^2\rho(r)/2)^{1/3}$,
$M=940 $ MeV the nucleon mass, and $E(\vec{p}\,)= \sqrt{M^2 +
\vec{p}^{\,2}}$. We will work on a non-symmetric nuclear matter with
different Fermi sea levels for protons, $k_F^p$, than for neutrons,
$k_F^n$ (equation above, but replacing $\rho /2$ by $\rho_p $ or
$\rho_n$, with $\rho=\rho_p + \rho_n$).  On the other hand, for the
$Z^0NN$ vertex we take ($N=n$ or $p$)
\begin{equation}
< N; \vec{p}^{~\prime}=\vec{p}+\vec{q}~ | j^\alpha_{nc}(0) | N;
\vec{p}~> =
\bar{u}(\vec{p}{~^\prime})(V^\alpha_N-A^\alpha_N)u(p) 
\end{equation}
with spinor normalization given by ${\bar u} u = 2m$, and vector and axial
 nucleon currents given by
\begin{equation}
V^\alpha_N = 2 \times  \left ( F_1^Z(q^2)\gamma^\alpha + i 
\mu_Z \frac{F_2^Z(q^2)}{2M}\sigma^{\alpha\nu}q_\nu\right)_N, \qquad
A^\alpha_N =  \left ( G_A^Z(q^2) \gamma^\alpha\gamma_5   + G_P^Z(q^2)
 q^\alpha\gamma_5 \right)_N
\end{equation}
Invariance under G-parity has been assumed to discard a term of the
form $(p^\mu+p^{\prime \mu})\gamma_5$ in the axial sector, while 
invariance under time reversal guarantees that all form
factors are real. Besides, and thanks to SU(3) symmetry, some relations 
exist among the NC form factors and the CC ($F_{1,2}^V$, $G_A$ and
$G_P$, see Ref.~\cite{NAV05}) and the electromagnetic ones\footnote{We use
the parameterization of Galster and collaborators~\protect\cite{Ga71},
which is also compiled  in Ref.~\protect\cite{NAV05}.} 
($F_1^{p,n}$, $\mu_p F_2^p$  and $\mu_n F_2^n$)
\begin{eqnarray}
\left(F_1^Z\right)^{p,n} &=& \pm F_1^V - 2\sin^2\theta_W F_1^{p,n}-
\frac12 F_1^s \\
\left(\mu_Z F_2^Z\right)^{p,n} &=& \pm \mu_V F_2^V - 2\sin^2\theta_W 
\mu_{p,n}F_2^{p,n}- \frac12 \mu_s F_2^s \\
\left(G_{A,P}^Z\right)^{p,n} &=& \pm G_{A,P} - G_{A,P}^s  
\end{eqnarray}
where $F_1^s, \mu_s F_2^s, G_A^s$ and $G_P^s$ are the strange vector
and axial nucleon form factors~\cite{Al96}. The pseudoscalar part of
the axial current does not contribute to the differential cross
section for massless neutrinos and for the rest of strange form
factors we use the results of the fit II of Ref.~\cite{Ga93},
\begin{equation}
G_A^s(q^2) = \frac{g_A^s}{(1-q^2/(M_A^s)^2)^2}, \quad F_1^s(q^2) =
\mu_s F_2^s(q^2) = 0
\end{equation}
with $g_A^s=-0.15$ and $M_A^s=1049$ MeV.
With all of these ingredients is straightforward to evaluate the 
contribution to the $Z^0-$selfenergy  of the
first  diagram of Fig.~\ref{fig:fig2}, which leads to 
\begin{eqnarray}
W^{\mu\nu}(q^0,\vec{q}\,) 
&=& - \frac{1}{2M^2} \int_0^\infty dr r^2 \Big \{
2 \Theta(q^0) \int \frac{d^3p}{(2\pi)^3}\frac{M}{E(\vec{p})}
\frac{M}{E(\vec{p}+\vec{q})} (-\pi) \nonumber\\  
&\times& \sum_{N=n,p} \Big \{ \Theta(k_F^N(r)-|\vec{p}~|)
\Theta(|\vec{p}+\vec{q}~|-k_F^N (r))
  A_N^{\nu\mu}(p,q)  \Big \}_{p^0=E(\vec{p})~}\delta(q^0 -Q+
E(\vec{p}) -E(\vec{p}+\vec{q}~)) \label{eq:res}
\end{eqnarray}
where $Q$ is the experimental $Q-$value, included in order to properly
reproduce the energy threshold. For inclusive observables we have set
it to zero to approximately take into account the possibility of
elastic scattering and transitions to nuclear excited states.  The
$d^3p$ integrations above can be analytically done and all of them are
determined by the imaginary part of the relativistic Lindhard
function, $\overline {U}_R(q,k_F^N,k_F^N)$.  Explicit expressions can
be found in Appendix B of Ref.~\cite{NAV05}. The NC nucleon tensor,
$A_N^{\nu\mu}$, can be found in Appendix~\ref{sec:app_nc}.  The
non-relativistic reduction of the hadronic tensor can be obtained by
replacing the factors $M/E(\vec{p})$ and $M/E(\vec{p}+\vec{q})$ in
Eq.~(\ref{eq:res}) by one. Explicit expressions can be found in
Appendix C of Ref.~\cite{NAV05}.

To finish  this section, we devote a few words to the Low Density
Theorem (LDT). At low nuclear densities the
imaginary part of the relativistic 
Lindhard function can be approximated by
\begin{equation}
{\rm Im}\overline{U}_R(q,k_F^N,k_F^N) \approx - \pi \rho_N
\frac{M}{E(\vec{q}\,)} \delta(q^0+M-E(\vec{q}\,))
\end{equation}
and thus one readily finds
\begin{equation}
\sigma_{\nu_l +\,
  A_Z \to \nu_l  + X } \approx  N \sigma_{\nu_l  +\, n \to \nu_l
   + n}+ Z \sigma_{\nu_l  +\, p \to \nu_l
   + p}, \qquad N=A-Z  \label{eq:ldt}
\end{equation}
which agrees with the LDT. 

\subsection{QE contribution: RPA nuclear correlation and FSI effects}
\label{sec:rpa}
Pauli blocking, through the imaginary part of the Lindhard function,
is the main nuclear effect included in the hadronic tensor of
Eq.~(\ref{eq:res}). RPA nuclear correlation and FSI effects played a
crucial role for inclusive QE CC neutrino-nucleus reactions, and we
include those here, following the same formalism as in Ref.~\cite{NAV05}.

\subsubsection{RPA}
\begin{figure}[b]
\centerline{\includegraphics[height=14.0cm]{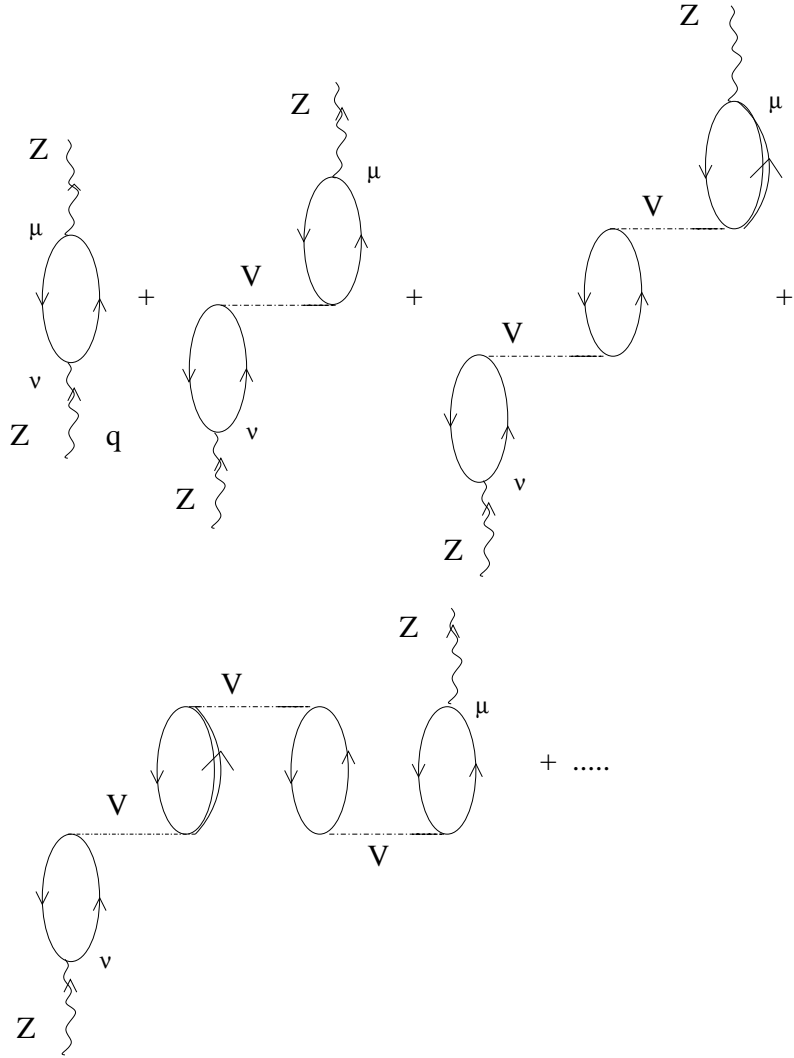}}
\caption{ Set of irreducible diagrams responsible for the
  polarization (RPA) effects in the 1p1h contribution to the
  $Z-$selfenergy. }\label{fig:fig3}
\end{figure}

We replace the 1p1h contribution to the $Z^0$ selfenergy by an
RPA response as shown diagrammatically in Fig.~\ref{fig:fig3}. For that
purpose we use an effective ph--ph interaction of the Landau-Migdal
type, with explicit contribution of pion and rho meson exchanges,
\begin{equation}
\begin{array}{ll}
V = & c_{0}\Big\{
f_{0}(\rho)+f_{0}^{\prime}(\rho)\vec{\tau}_{1}\vec{\tau}_{2}+
g_{0}(\rho)\vec{\sigma}_{1}\vec{\sigma}_{2} \Big\}+ 
\vec{\tau}_{1}\vec{\tau}_{2} \Big 
\{V_l(q) \vec{\sigma}_{1}\hat{q}\vec{\sigma}_{2}\hat{q}+ V_t(q) (
\vec{\sigma}_{1}\vec{\sigma}_{2} -
\vec{\sigma}_{1}\hat{q}\vec{\sigma}_{2}\hat{q}) \Big \}
\end{array} \label{eq:lm}
\end{equation}
where $\vec{\sigma}$ and $\vec{\tau}$ are Pauli matrices acting on the
nucleon spin and isospin spaces, respectively, and $\hat{q} =
\vec{q}/|\vec{q}\,|$.  We take the coefficients $f_0$, $f^\prime_0$ and $g_0$
from Ref.~\cite{Sp77},
\begin{equation}
f_{i}(\rho (r))=\frac{\rho (r)}{\rho (0)} f_{i}^{(in)}+
\left[ 1-\frac{\rho (r)}{\rho (0)}\right] f_{i}^{(ex)}
\end{equation}
where
\begin{equation}
\begin{array}{ll}
f_{0}^{(in)}=0.07 & f_{0}^{\prime (ex)}=0.45\\ 
f_{0}^{(ex)}=-2.15 &
f_{0}^{\prime (in)}= 0.33 \\ 
g_{0}^{(in)}=g_{0}^{(ex)}=g_{0}=0.575 ~~~~~~~~~& \\
\end{array}
\end{equation}
and $c_{0}=380\, {\rm MeV fm}^{3}$. In the $S = 1 = T$ channel
($\vec{\sigma} \vec{\sigma} \vec{\tau} \vec{\tau}$ operator) we use an
interaction with explicit $\pi$ (longitudinal) and $\rho$ (transverse)
exchanges, which has been used for the renormalization of the pionic
and pion related channels in different nuclear reactions at
intermediate energies~\cite{pion}--\cite{CO92}. The strengths of the ph-ph
interaction in the longitudinal and transverse channel are given by
\begin{eqnarray}
V_l(q^0,\vec{q}) &=& \frac{f^2}{m^2_\pi}\left
\{\left(\frac{\Lambda_\pi^2-m_\pi^2}{\Lambda_\pi^2-q^2 }\right)^2
\frac{\vec{q}{\,^2}}{q^2-m_\pi^2} + g^\prime_l(q)\right \}, \qquad
\frac{f^2}{4\pi}=0.08,~~\Lambda_\pi=1200~{\rm MeV} \nonumber \\
V_t(q^0,\vec{q}) &=& \frac{f^2}{m^2_\pi}\left
\{   C_\rho \left (\frac{\Lambda_\rho^2-m_\rho^2}{\Lambda_\rho^2-q^2 }\right)^2
\frac{\vec{q}{\,^2}}{q^2-m_\rho^2} + g^\prime_t(q)\right \}, \qquad
C_\rho=2,~~\Lambda_\rho=2500~{\rm MeV},~~m_\rho=770~{\rm MeV} \label{eq:st2}
\end{eqnarray}
The SRC functions $g^\prime_l$ and $g^\prime_t$ have a smooth
$q-$dependence~\cite{OTW82,Ga88}, which we will not consider
here\footnote{This is justified because taking into account the
$q-$dependence leads to minor changes for low and intermediate
energies and momenta, where this effective ph-ph interaction should be
used.}, and thus we will take
$g^\prime_l(q)=g^\prime_t(q)=g^\prime=0.63$ as it was done in the
study of inclusive nuclear electron scattering carried out in
Ref.~\cite{GNO97}, and also in some of the works of
Ref.~\cite{pion,pion1}. 

The above interaction has been successfully
tested in different nuclear processes at intermediate
energies~\cite{OTW82}--\cite{GNO97bis}, and we
will use here the same form and parameters as in our recent study of
CC neutrino nucleus reactions~\cite{NAV05}. As it is explained there,
$\Delta(1232)$ degrees of freedom are also taken into account. Given
the spin-isospin quantum numbers of the $\Delta$ resonance, these
degrees of freedom only modify the vector-isovector ($S = 1 = T$)
channel of the RPA response function. The ph--$\Delta$h and
$\Delta$h--$\Delta$h effective interactions are obtained from the
interaction of Eq.~(\ref{eq:lm}) by replacing $\vec{\sigma} \to
\vec{S} $, $\vec{\tau} \to \vec{T} $, where $\vec{S},\vec{T} $ are the
spin, isospin $N\Delta$ transition operators~\cite{OTW82} and $f\to
f^*=2.13~f$, for any $\Delta$ which replaces a nucleon.

Thus, the $V$ lines in Fig.~\ref{fig:fig3} stand for the effective
ph($\Delta$h)-ph($\Delta$h) interaction described so far.  Keeping
track of the  operators responsible, we have examined and renormalized
all different contributions to the NC nucleon tensor $A_N^{\mu\nu}$, by
summing up the RPA series depicted in Fig.~\ref{fig:fig3}. The
procedure is discussed in detail in Ref.~\cite{NAV05}, now however the
isospin structure of the $Z^0NN$ vertex 
\begin{equation}
\Gamma^\mu \Big (F_p \frac{1+\tau_z}{2} + F_n \frac{1-\tau_z}{2}\Big)
\end{equation}
with $\Gamma^\mu$ some matrix in the Dirac space and $F$ some form
factor, does not select only the isovector channels of the in medium
effective ph($\Delta$h)-ph($\Delta$h) interaction, as it is the case
for the CC induced processes studied in~\cite{NAV05}, and hence we
also find contributions from the isoscalar part of the effective
force.  

For simplicity and to compute these polarization corrections, we have
assumed a symmetric nuclear matter with the same Fermi sea level for
protons and neutrons. On the other hand, since the
ph($\Delta$h)-ph($\Delta$h) effective interaction is non-relativistic,
we have computed polarization effects only for the leading and
next-to-leading terms in the $p/M$ expansion. Thus, order ${\cal
O}\left(k_F\vec{p}^{\,2}/M^2,k_F\vec{p}^{\, \prime
2}/M^2,k_Fq^0/M\right)$ has been neglected. We have made an exception
to the above rule, and since $\mu_Z$ could be relatively large, we
have taken $\mu_Z F_2^Z|\vec{q}\,|/M$ to be of order ${\cal O}(0)$ in
the $p/M$ expansion.  With all these ingredients, we find $\delta 
W^{\mu\nu}_{\rm RPA}$,  contribution which  has to be added to
the hadronic tensor, $W^{\mu\nu}$, given in Eq.~(\ref{eq:res}) to
account for the medium polarization effects,
\begin{eqnarray}
\delta W^{\mu\nu}_{\rm RPA} &=& - \frac{1}{2M^2} \int_0^\infty dr r^2 \Big \{
2 \Theta(q^0) \int \frac{d^3p}{(2\pi)^3}\frac{M}{E(\vec{p})}
\frac{M}{E(\vec{p}+\vec{q})} (-\pi)\delta(q^0 +
E(\vec{p}) -E(\vec{p}+\vec{q}~)) \nonumber\\  
&\times&  \Big \{ \Theta(k_F(r)-|\vec{p}~|)
\Theta(|\vec{p}+\vec{q}~|-k_F (r))
  \delta A^{\nu\mu}_{\rm RPA}(p,q)  \Big \}_{p^0=E(\vec{p})~}\label{eq:rpa}
\end{eqnarray}
The $00,0z,zz,xx$ and $xy$ components of the RPA contribution to the
NC nucleon tensor, $\delta A^{\nu\mu}_{\rm RPA}(p,q)$, are given in
Sect.~\ref{sec:rpaamunu} of the Appendix. There, the density dependent
tensor $\delta A^{\nu\mu}_{\rm RPA}(p,q)$ is given in terms of the parameters
of the effective ph($\Delta$h)-ph($\Delta$h) force and
$U(q,k_F)=U_N+U_\Delta$, the Lindhard function of Ref.~\cite{Ga88},
which for simplicity we evaluate\footnote{The functions $U_N$ and
$U_\Delta$ are defined in Eqs.(2.9) and (3.4) of
Ref.~\protect\cite{Ga88}, respectively.} in an isospin symmetric
nuclear medium of density $\rho$. The different couplings for $N$ and
$\Delta$ are incorporated in $U_N$ and $U_\Delta$ and then the same
interaction strengths $V_l$ and $V_t$ are used for ph and $\Delta $h
excitations~\cite{OTW82}--\cite{GNO97bis}. Note that, backward (crossed term of
the Lindhard function) propagating ph and $\Delta h$ excitations are
also taken into account within our framework. For
positive values of $q^0$ the backward propagating ph excitation has no
imaginary part, and for QE kinematics $U_\Delta$ is also real.

Finally, we should stress that the $f^\prime_0$, $f_0$ and $g_0$ terms
of the effective interaction cannot produce $\Delta$h excitations and
therefore, when these terms are involved in the RPA renormalization,
only the nucleon Lindhard function ($U_N$) appears (see
Eqs.~(\ref{eq:coeffs})).

\subsubsection{FSI: dressed nucleon propagators  in
the nuclear medium}
\label{sec:fsi}

Once a ph excitation is produced by the virtual $Z^0-$boson, the
nucleon is interacting with the rest of nucleons of the nucleus,
colliding  many times, thus inducing the emission of
other nucleons. The result of it is a quenching of the QE peak respect
to the simple ph excitation calculation and a spreading of the
strength, or widening of the peak. In our many body scheme we will
account for the FSI by using nucleon propagators properly dressed with
a realistic selfenergy in the medium, which depends explicitly on the
energy and the momentum~\cite{FO92}.  Hence we substitute the particle
nucleon propagator, $G(p;\rho)$, in Eq.~(\ref{eq:Gp}) by a
renormalized nucleon propagator, $G_{\rm FSI}(p;\rho)$, including the
nucleon selfenergy in the medium, $\Sigma (p^0 , \vec{p}\,; \rho)$,
\begin{equation}
 G_{\rm FSI}(p;  \rho)= \frac{1}{p^{0}-{\bar E}(\vec{p}\,)-\Sigma(p^{0},
  \vec{p}\,;\rho)} \label{eq:gfsi}
\end{equation}
with ${\bar E}(\vec{p}\,) = M + \vec{p}^{\,2}/2M$. This approach led
to excellent results in the study of inclusive electron and CC
neutrino scattering from nuclei~\cite{GNO97,NAV05}. Since the model of
Ref.~\cite{FO92} is not Lorentz relativistic and it also considers an
isospin symmetric nuclear medium, we will only discuss the FSI effects
for nuclei with approximately equal number of protons and neutrons,
and using non-relativistic kinematics for the nucleons. Thus, we have obtained
Eq.~(\ref{eq:gfsi}) from the non-relativistic reduction of
$G(p;\rho)$, in Eq.~(\ref{eq:Gp}), by including the nucleon selfenergy.

 To account for FSI effects in an isospin symmetric
nuclear medium of density $\rho$ we should make the following
substitution~\cite{NAV05,GNO97}
\begin{eqnarray}
& & 2 \Theta(q^0) \int \frac{d^3p}{(2\pi)^3}
\Theta(k_F(r)-|\vec{p}~|) \Theta(|\vec{p}+\vec{q}~|-k_F(r))
   (-\pi)\delta(q^0 +
{\bar E}(\vec{p}) -{\bar E}(\vec{p}+\vec{q}~))
{\cal A}^{\nu\mu}(p,q)|_{p^0={\bar E}(\vec{p})~} \nonumber \\
&\rightarrow&  -\frac{\Theta(q^0)}{4\pi^2}\int d^3p 
 \int_{\mu-q^0}^\mu d\omega S_h(\omega,\vec{p}\,;\rho)
S_p(q^0+\omega,\vec{p}+\vec{q}\,;\rho) 
{\cal A}^{\nu\mu}(p,q)|_{p^0={\bar E}(\vec{p})~},\quad {\cal
  A}^{\mu\nu}=A_N^{\mu\nu}, \delta A^{\mu\nu}_{\rm RPA} \label{eq:fsi}
\end{eqnarray}
in the expression of the hadronic tensor (Eqs.~(\ref{eq:res})
and~(\ref{eq:rpa}) ). $S_h, S_p$ are the hole and particle spectral
functions related to nucleon selfenergy $\Sigma$ by means of
\begin{eqnarray}
\label{eq:spec}
S_{p,h}(\omega,\vec{p}\,;\rho) &=& \mp\frac{1}{\pi}\frac{{\rm
Im}\Sigma(\omega,\vec{p}\,;\rho)}{\Big[\omega-{\bar E}(\vec{p}\,)-{\rm
Re}\Sigma(\omega,\vec{p}\,;\rho)  \Big]^2+\Big[{\rm
Im}\Sigma(\omega,\vec{p}\,;\rho)\Big]^2}
\end{eqnarray}
with $\omega\ge \mu$ or $\omega\le \mu$  for $S_p$ and $S_h$,
respectively.  The chemical potential $\mu$ is determined by
\begin{equation}
\mu = M + \frac{k_F^2}{2M} + {\rm Re}\Sigma(\mu, k_F)
\end{equation}

 The $d^3p$ integrations have to be done now numerically, and since the
imaginary part of the nucleon selfenergy for the hole states is much
smaller than that of the particle states at intermediate nuclear
excitation energies, we make the approximation of setting to zero
Im$\Sigma$ for the hole states. Thus, we take
\begin{equation}
S_{h}(\omega,\vec{p}\,;\rho)=\delta(\omega-\hat{E}(\vec{p}\,))
\Theta(\mu-\hat{E}(p))
\end{equation}
where $\hat{E} (p)$ is the energy associated to a momentum $\vec{p}$
obtained self consistently by means of the equation
\begin{equation}
\hat{E}(\vec{p}\,)={\bar E}(\vec{p}\,) + {\rm Re} \Sigma (\hat{E}(\vec{p}\,),
    \vec{p}\,;\rho)
\end{equation}
The same approximation is also used in some of the calculations for the 
particle spectral function. The effects of this approximation will be discussed
in Sect. \ref{sec:exclu}.

\section{The Monte Carlo simulation}
\label{sec:MC}

\subsection{Kinematics of the outgoing nucleon in the first step}

In a previous work~\cite{NAV05} and in Sect.~\ref{sec:nc} we carried
out a thorough evaluation of the CC and NC inclusive neutrino and
antineutrino induced nuclear reactions in the QE region. Thus we have
determined, for a fixed incoming neutrino or antineutrino LAB energy,
the inclusive QE cross section $d^2\sigma /d\Omega^{\prime}
dE^{\prime}$ ($\Omega^{\prime}$, $E^{\prime}$ are the solid angle and
energy of the outgoing lepton).  The absorption of the gauge boson
($W^\pm$ or $Z^0$) with four momentum $q^\mu$ by one nucleon
constitutes the reaction mechanism (1p1h excitation)\footnote{ This is
not entirely correct, since two nucleon absorption modes are also
considered, when the FSI effects described in Sect.\ref{sec:fsi} are
taken into account. We will discuss this point later in
Subsect.~\ref{sec:exclu}.}, and the corresponding reaction probability
is determined by $d^2\sigma /d\Omega^{\prime} dE^{\prime}$. Moreover,
within our scheme, we obtain $d^2\sigma /d\Omega^{\prime} dE^{\prime}$
after performing an integration over the whole nuclear volume (see
Eqs.~(16) and (\ref{eq:res}) of Ref.~\cite{NAV05} and
Sect.~\ref{sec:nc}, respectively). Thus, for a fixed transferred four
momentum $q^\mu$, chosen according to $d^2\sigma /d\Omega^{\prime}
dE^{\prime}$, we can randomly select the point of the nucleus where
the absorption takes place using the profile $d^5\sigma
/d\Omega^{\prime} dE^{\prime}d^3r$ evaluated in Ref.~\cite{NAV05} for
CC and in Sect.~\ref{sec:nc} for NC.

Now, we need on top of that to have the distribution of three--momenta
of the outgoing nucleon. This can be done by not performing the
integration over the three momentum, $\vec{p}$, of the nucleon
occupied states in ~\cite{NAV05} and in Sect.~\ref{sec:nc} (see for
instance Eq.~(\ref{eq:res}) for the the NC case), since the outgoing
nucleon momentum is $\vec{p}+\vec{q}$.  However, in order to
incorporate the nucleon scattering with other nucleons, we choose the
alternative procedure of generating events probabilistically one by
one, with a weight given by $d^5\sigma /d\Omega^{\prime}
dE^{\prime}d^3r$.  For each event in a certain position $\vec{r}$ and
with a transferred four momentum $q^\mu$ we generate a random momentum
$\vec{p}$ from the local Fermi sea. The vector $\vec{p}+\vec{q}$ gives
us the direction of the nucleon.  The energy of the nucleon is then
obtained by imposing energy conservation assuming in this step and
throughout all the MC simulation that the nucleons move
in a Fermi gas under an attractive potential equal to the local Fermi
energy and therefore
\begin{equation}
     \tilde{E}'=\tilde{E}(\vec{p})+q^0
\end{equation}
where $\tilde{E}(\vec{p})=\sqrt{\vec{p}^2+M^2}-k_F^2(r)/2M$.  This
will provides us with the modulus of the outgoing
momentum\footnote{Indeed, we have $|{\vec{p}\,}^{\prime}|^2 =
(\tilde{E}' + k_F^2(r)/2M)^2 - M^2$.}  $\vec{p}^{\,\prime}$.  If it
happens that $|{\vec{p}\,}^{\prime}|<k_{F}(r)$ (the local Fermi
momentum) then the event is Pauli blocked, it is dismissed and another
event is generated. Thus, we have already the configuration of the
final state after the first step, namely, one nucleon produced in the
point $\vec{r}$ of the nucleus with momentum ${\vec{p}\,}^{\prime}$.
With respect to having a proton or a neutron in the final state,
this is trivially done: for the CC case, the outgoing nucleon is a
proton (neutron) for neutrino (antineutrino) induced processes, while
for NC, in Sect.~\ref{sec:nc}, the reaction probability was already
split into a proton and a neutron induced ones (see for instance
Eq.~(\ref{eq:res}))\footnote{Note that polarization corrections
(Eq.~(\ref{eq:rpa})) were estimated in isospin symmetric nuclear
matter, and thus approximately the RPA term equally contributes to
both the proton and neutron $Z^0$ absorption channels.}.

\subsection{Nucleon propagation}

The nucleons in the nucleus move under the influence of a
complex optical potential. The imaginary part of the potential
is related to the probability of nucleon quasielastic
collisions in the nucleus (and extra pion production at higher
energies, which we do not consider here). We consider explicitly
these collisions since they generate new nucleons going outside
the nucleus. As with respect to the real part, we use it to
determine the classical trajectories that the nucleons follow
in the nucleus between collisions.
       
As done in Refs.~\cite{GNO97bis,COV94} we take as the real part
of the nucleon-nucleus potential
\begin{equation}
     V(r)=V_{\infty}-{\cal{E}} (r)=-\frac{k_{F}^{2}}{2M}=
     -\frac{1}{2M} \left(
           \frac{3}{2}\pi^{2}\rho (r)\right)^{2/3}
\end{equation}
It represents the interaction of a single nucleon with the average
potential due to the rest of the nucleons. This choice of $V(r)$ means
that the total nucleon energy is the difference between its kinetic
energy and the Fermi energy, $k_{F}^{2}/2M$.

With respect to collisions in our MC simulation we follow
each excited nucleon by letting it move a short distance $d$ such that
$Pd<<1$ ($P$ represents the probability per unit of length for a
quasielastic collision ). The new position (${\vec{r}\,}^{\prime}$)
and momentum (${\vec{p}\,}^{\prime}$) are taken from the Hamiltonian
equations as
\begin{eqnarray}
{\vec{r}\,}^{\prime}&=&\vec{r}+\delta\vec{r}=\vec{r}+
\frac{\vec{p}}{|\vec{p}\,|}d \nonumber\\
{\vec{p}\,}^{\prime}&=&\vec{p}+\delta\vec{p} \,\,;\,\, 
           \delta\vec {p}\,= - \frac{\partial V}{\partial r} 
           \frac{E(p) d}{p} \frac{\vec{r}}{|\vec{r}\,|}
\end{eqnarray}
which follow from the Hamilton equations or equivalently from energy 
and angular momentum conservation.
                        
Our code selects randomly, according to the reaction probabilities
which will be discussed in Subsect.~\ref{sec:nn}, if the nucleon is
scattered or not and, in the case of scattering, what kind of process
takes place.  If no collision takes place, we move the nucleon again.
When the nucleon leaves the nucleus we stop the process and it is
counted as a contribution to the cross section. If a $NN$ scattering
is selected instead, we take a random nucleon from the Fermi sea and
calculate the initial kinematical variables ($P^{\mu}$ and $s$, full
four-momentum of the nucleon-nucleon system in the nuclear frame and
invariant energy, respectively). Then, a $\cos\theta_{c.m.}^{N_2}$ is
selected, according to the expression given in Eq.~(\ref{eq:secnn})
below and taken from Ref.~\cite{COV94}.  This expression gives us the
correct probability given by $d\sigma^{NN}/d\Omega_{c.m.}$ plus
Pauli blocking restrictions. We take also into account Fermi motion
and renormalization effects in the angular dependence. We take them
into account by multiplying each event by a weight factor
\begin{equation}
\xi =\lambda (N_{1},N_{2})\hat{\sigma}^{N_{1}N_{2}}\rho_{N_{2}}
\end{equation}
where $1/\lambda(N_{1},N_{2})$ is the probability by unit of length
that one nucleon $N_1$ collides with another nucleon $N_2$. In
average, this factor $\xi$ is equal to one. Explicit formulas are
given in the next subsection.

Our method assumes that the nucleons propagate semiclassically in the nucleus.
The justification of this hypothesis for reactions induced by real photons is
given in Ref.~\cite{COV94}, and has also been successfully used in 
Refs.~\cite{Ramos:1996ik,VicenteVacas:1993bk}. In the next section we give some
detail on the evaluation of the equivalent $NN$ cross section in the medium.

\subsection{NN cross sections}
\label{sec:nn}       

We are using the parameterization of the $NN$ elastic cross section
given in the Appendix of Ref.~\cite{COV94}. Since for particles of low
momenta, the MC induces large errors, we are not considering
collisions of nucleons with kinetic energies below 30 MeV. This is to
say, we do not follow the path through the nuclear medium of nucleons with
kinetic energies below 30 MeV, and we just consider that those
nucleons get out of the  nucleus without suffering further collisions,
in which eventually they could change charge or loss some more energy.
          
On the other hand, the reaction probability will change due to the
nuclear medium effects (Fermi motion, Pauli blocking and medium
renormalization). Then, according to Ref.~\cite{COV94}, the expression
for the mean free path ($\lambda$) of the nucleon is given by
\begin{equation}
\frac{1}{\lambda(N_1)}
= 4 {\displaystyle \int}\frac{d^{3}p_{2}}{(2\pi)^{3}}
\left[
\Theta(k_F^p(r)-|\vec{p}_2|)\frac{Z}{A}\hat{\sigma}^{N_{1}p}(s)+
\Theta(k_F^n(r)-|\vec{p}_2|)\frac{(A\,-\,Z)}{A}\hat{\sigma}^{N_{1}n}(s)\right]
\frac{|\vec{p}_{1\,{\rm lab}}|}{|\vec{p}_{1}|}
\end{equation}
with $P^\mu$, the full four-momentum of the NN system in the nuclear
frame ($s=P^{2}=(p_1+p_2)^{2}$), $N_1$, the incoming nucleon and
$N_2$, the nucleon in the medium. The factor
$|\vec{p}_{1\,{\rm lab}}|/|\vec{p}_1|$ is related to the different flux of
particles in the nuclear frame and in the nucleon frame
($\vec{p}_{1\,{\rm lab}}$ is the incoming LAB momentum in the $NN$
system and $\vec{p}_1$, the momentum in the nuclear system). Furthermore,
\begin{equation}
\hat{\sigma}^{N_{1}N_{2}} = {\displaystyle \int}d\Omega_{\rm
c.m.}\frac{d\sigma^{N_{1}N_{2}}} {d\Omega_{\rm c.m.}}{\bf C_T(q,\rho)}
\Theta\Big(\kappa-\frac{|\vec{P}\cdot\,\vec{p}_{\rm
c.m.}|}{|\vec{P}\,||\vec{p}_{c.m.}|}\Big) \label{eq:secnn}
\end{equation}
where c.m. is the $NN$ center of mass frame and
\begin{equation}
  \kappa = x\Theta(1-|x|)\,+\,\frac{x}{|x|}\Theta(|x|-1), 
\quad x =   \frac{P^{0}\,p^{0}_{c.m.}-\epsilon_{F}\sqrt{s}}
   {|\vec{P}\,||\vec{p}_{c.m.}|}
\end{equation}
where $\vec{p}_{c.m.}$ is the nucleon momentum in the c.m. frame,
$\epsilon_{F}$ is the Fermi energy ($k_{F}^{2}/2M$), and the density
dependent factor ${\bf C_T}$ is defined in Eq.~(\ref{eq:coeffs}), with
$q$ the momentum transfer in the nuclear frame. In these expressions,
$\Theta(\kappa-|\hat{P}.\,\hat{p}_{\rm c.m.}|)$ takes into account Pauli
blocking and ${\bf C_T(q,\rho)}$, the nuclear medium renormalization.

\section{Results}
\label{sec:RS}

We compile in Table \ref{tab:nuclei} the data used for the nuclei studied in
this work. As in Ref.~\cite{NAV05}, nuclear masses and charge densities are
taken from Refs.~\cite{Fi96} and~\cite{Ja74}, respectively. The neutron density
is taken with the same form as the charge density but properly normalized and
with a different radius as suggested by Hartree-Fock calculations~\cite{Ne75}
and corroborated by pionic atom data~\cite{GNO92}. Furthermore, charge
densities  do not correspond to proton point--like densities because of the
finite size of the nucleon. This is taken  into account by following the
procedure outlined in Section II of Ref.~\cite{GNO92}.

In the case of NC driven processes, the minimum energy transfer,
$q^0$, needed for a proton (neutron) emission reaction corresponds to
the proton (neutron) separation energy $Q_p$ ($Q_n$). Our Fermi gas
model does not account properly for this minimum
energy~\footnote{Indeed, in a Fermi model we need zero excitation
energy to emit nucleons.} and we correct it by replacing 
\begin{equation}
 q^0 \rightarrow q^0- Q_p(A_Z)
\end{equation}
for the proton emission reaction and similarly for the neutron
one. In the case of CC processes, this minimum excitation energy is
$Q+ Q_p(A_{Z+1})$ $\left [\, \overline{Q} + Q_n (A_{Z-1})\right]$  for neutrino
[antineutrino] induced reactions, with $ Q=M(A_{Z+1})-M(A_Z)$
$\left[\,\overline{Q}=M(A_{Z-1})-M(A_Z)\right]$. 

In what follows, we will present different results for NC induced
inclusive and NC and CC nucleon emission reactions at moderate energy
transfers to the nucleus, which illustrate the important role played by
the different nuclear effects considered in this work: Pauli blocking,
RPA and FSI effects, and finally the re-scattering of the outgoing
nucleon. For consistency we will always use  non-relativistic
kinematics to evaluate the contribution of a particle--hole
excitation. Relativistic effects were studied in Ref.~\cite{NAV05} and
found there to be small at the energy regime studied here.

All differential cross sections shown in this section are computed in
the LAB frame.

\begin{table}
\begin{center} \begin{tabular}{ccccc|cc|cccc}\hline\tstrut 
Nucleus & $R_{p}$ & $R_{n} $ & $a_p^*$ & $a_n^*$ 
        & $Q_p(A_Z)$     &  $Q_n(A_Z)$  
        & $Q_p(A_{Z+1})$ &  $Q_n(A_{Z-1})$ 
        & $Q$ & $\overline{Q}$   \\\hline \tstrut
$^{16}$O&1.833&1.815&1.544&1.529&12.127&15.663&$-$
0.536&2.489&14.906 & 10.931 \\ 
$^{40}$Ar&3.47&3.64&0.569 & 0.569  & 12.528 & 9.869&7.582&5.830 &0.994&7.991\\
$^{40}$Ca&3.51&3.43&0.563&0.563&8.328&15.641&0.539&7.799&13.809&1.822 \\  
$^{208}$Pb&6.624&6.890&0.549&0.549&8.008&9.001&3.707&3.790&2.368&5.512\\  
\hline \end{tabular}
\\[1mm] (*) The parameter $a$ is dimensionless for the
 MHO density form.
 \end{center} 
\caption{ 
 Charge ($R_p, a_p$), neutron matter ($R_n, a_n$) density
parameters (in fermi units),
 and $Q_{p(n)}$ proton (neutron) energy separation, $Q-$ and $\overline{Q}-$
values for different nuclei in MeV units.  For the oxygen 
we  use a modified harmonic
 oscillator (MHO) density, $\rho(r) = \rho_0 (1+a(r/R)^2)\exp(-(r/R)^2)$,
 while for the rest of the nuclei, a two-parameter Fermi
 distribution, $\rho(r) = \rho_0  /(1+\exp((r-R)/a))$, was used. }  
\label{tab:nuclei} 
\end{table}

\subsection{Inclusive NC  scattering at low energies}
\label{sec:RSNC}
In this section we present results for the inclusive QE NC cross sections
for both neutrino and antineutrino beams in several nuclei
concentrating specially on the nuclear medium effects. The results
obtained with the same model for CC processes can be found in
Ref.~\cite{NAV05}.

\begin{figure}[tbh]
\begin{center}
\includegraphics[scale=0.7 , bb= 50 560 540 790]{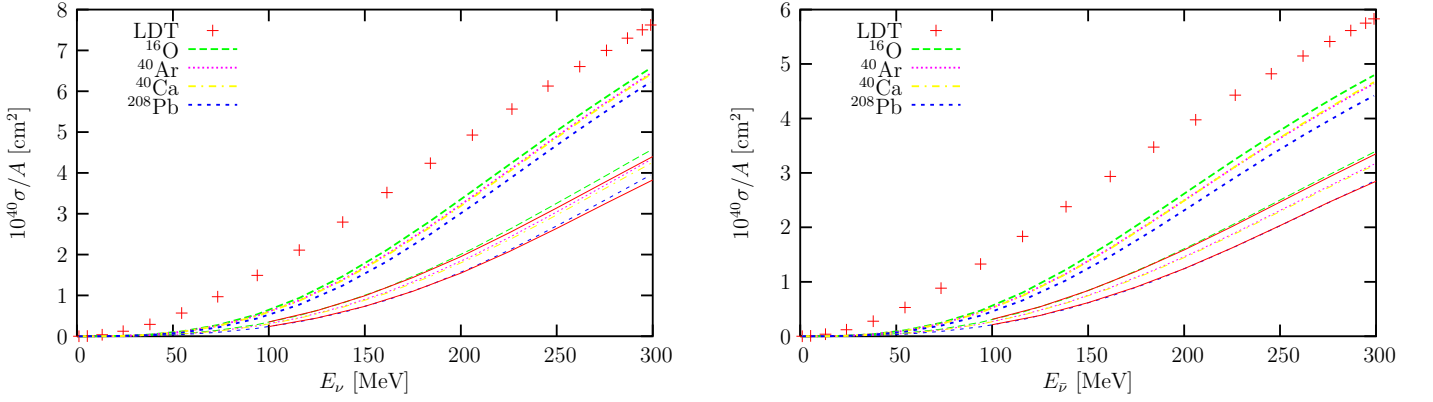}
\caption{ (color online). Cross section for the 
$\nu (\bar{\nu})+A\to \nu (\bar{\nu})+X$ processes as a function of the beam
energy. Crosses: $(\sigma_{\nu (\bar{\nu})+p\to \nu (\bar{\nu})+p}+
\sigma_{\nu (\bar{\nu})+n\to \nu (\bar{\nu})+n})/2$. Intermediate band: Cross
sections for several nuclei including Pauli blocking. Lower band: Cross
sections for several nuclei with Pauli blocking + RPA correlations. 
The two solid lines of the lower band correspond to the full model calculation
(including also FSI) for oxygen and lead.
The left and right panel correspond to neutrino and antineutrino
beam respectively.}
\label{fig:res1}
\end{center}
\end{figure}

In Fig.~\ref{fig:res1} we show the cross section for the processes
$\nu (\bar{\nu})+A\to \nu (\bar{\nu})+X$ at low and intermediate
energies. Above the shown range, pion production could become
relevant. We also show for comparison the isospin averaged neutrino
(antineutrino) free nucleon cross section.

We find that nuclear effects produce a strong reduction of the cross section. 
The simple consideration of Pauli blocking is already quite effective
even at the higher studied energies. The inclusion of the long range
RPA correlations is also quite significant. On the other hand, the
magnitude of the reduction is only weakly dependent on the isospin and
the atomic mass showing, as expected, a larger effect for heavier
nuclei. The size of the effect is similar to the one found in
Ref.~\cite{NAV05} for CC processes although in this case we also
explore the isoscalar pieces of the effective ph-ph interaction
responsible of the RPA correlations and the energy thresholds are different.

The addition of FSI (Sect. \ref{sec:fsi}) to the calculation
containing already Pauli blocking and RPA correlations does not affect
practically the results for the totally integrated cross sections.

\begin{figure}[tbh]
\centerline{\includegraphics[scale=0.7, bb= 50 560 540 790]{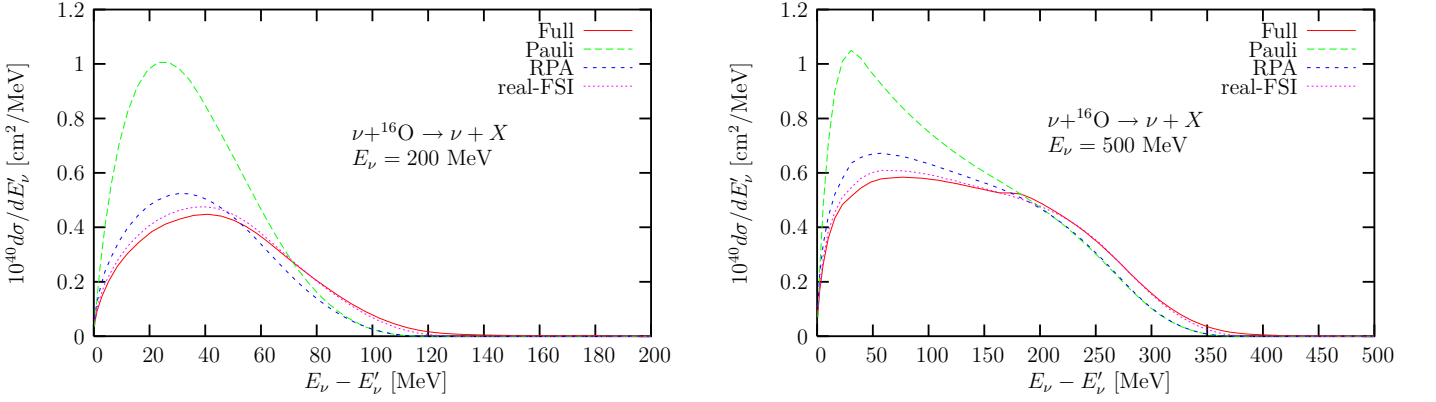}}
\caption{  (color online). NC neutrino $d \sigma/d
   E_{\nu'}$ cross section as a function of the energy transfer and
   for two different incoming neutrino energies. Predictions from
   different stages of refinements of the model are shown.  Long
   dashed line: Pauli blocking.  Short dashed line: Pauli blocking +
   RPA.  Dotted line: Pauli + RPA + FSI considering only the real part
   of the nucleon self-energy as explained in the text. Solid line:
   full model.}\label{fig:res2}
\end{figure}

We show in Fig.~\ref{fig:res2} the final neutrino energy spectrum for
two typical cases. Very similar results are obtained for other nuclei
and energies. Although this cross section is not experimentally
observable it is useful to show how nuclear effects modify the
spectrum and whether they favor some energy transfer.
We find that RPA produces a quite smooth reduction that covers all the
energy range, although the effects produced by the RPA correlations
are weaker when the energy transferred to the nucleus is larger, much
the same as it is the case for CC processes.  The inclusion of FSI
spreads the spectrum allowing for larger energy transfers.

\begin{figure}[tbh]
\begin{center}
\includegraphics[scale=0.9, bb= 50 560 540 790]{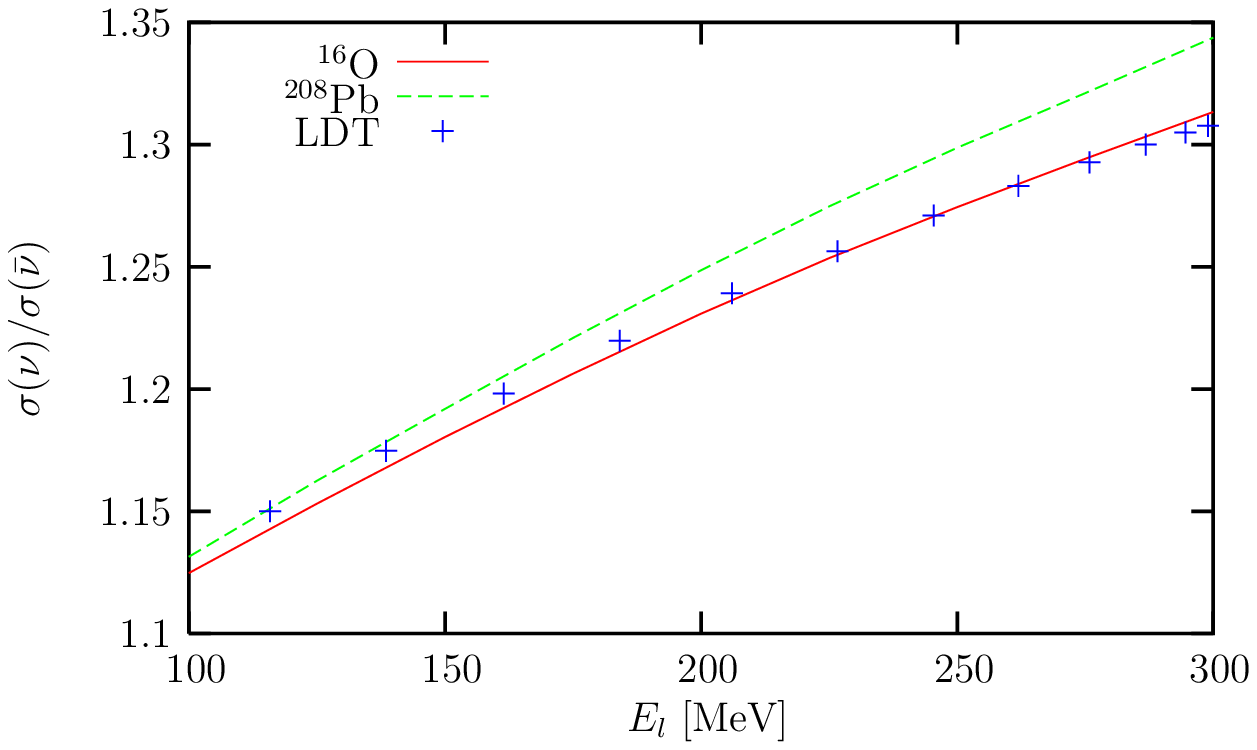}
\caption{  (color online). 
$\sigma_{\nu}/\sigma_{\bar{\nu}}$ ratio 
for the $\nu(\bar{\nu})+A\to \nu'(\bar{\nu}')+X$ reaction as a function
of the energy for several nuclei. LDT means free nucleon results.
In all other lines the full model is used.}\label{fig:res3}
\end{center}
\end{figure}
In Fig.~\ref{fig:res3}, we show the ratio of the cross sections
induced by neutrino and antineutrinos in oxygen and lead from our full
model and compare to the results obtained in the free case: LDT
(Eq.~(\ref{eq:ldt})).  The effect of the nuclear medium in the ratio
is quite small although it increases with energy and with the nuclear
mass. Similar results were found in Ref.~\cite{NAV05} for the CC
case. 

\subsection{Inclusive $(\nu_l,\nu_l N)$, $(\nu_l,l^- N)$, $({\bar \nu}_l,{\bar
\nu}_l N)$ and $({\bar \nu}_l,l^+ N)$ reactions in nuclei at low
energies}
\label{sec:exclu}

In this section we present results for  processes, in which the final
nucleons are  detected for both neutrino and antineutrino beams in several 
nuclei. We will also include CC reactions using for this case
the model of Ref.~\cite{NAV05} in which we have have implemented the nucleons
rescattering as described in Sec.~\ref{sec:MC}. 

As it is illustrated in Fig.~\ref{fig:res2}, the consideration of the 
dressing of the nucleon propagators (FSI)
produces a quenching of the QE peak respect to the simple ph
excitation calculation and a spreading of the strength, or widening of
the peak.  Most of the effect of the FSI comes from the consideration
of the real part of the nucleon selfenergy of Eq.~(\ref{eq:spec}) as
it is shown in the results obtained taking the limit of vanishing
${\rm Im}\Sigma$ (dotted curve in Fig.~\ref{fig:res2}). Thus, we find that
the change in the nucleon dispersion relation is more important than
the inclusion of the small nucleon width in the medium, related to the
quasielastic channels, which will account for $Z^0$ absorption by two
nucleons. Given the quality of this approximation for the production
of the spectra needed to start the MC simulation of the nucleons
rescattering, and the considerable reduction in time for its
evaluation, it has been used for all the MC results presented in this
work.

\subsubsection{CC Nucleon spectra} 

\begin{figure}[tbh]
\begin{center}
\includegraphics[scale=0.7, bb=50 350 540 790]{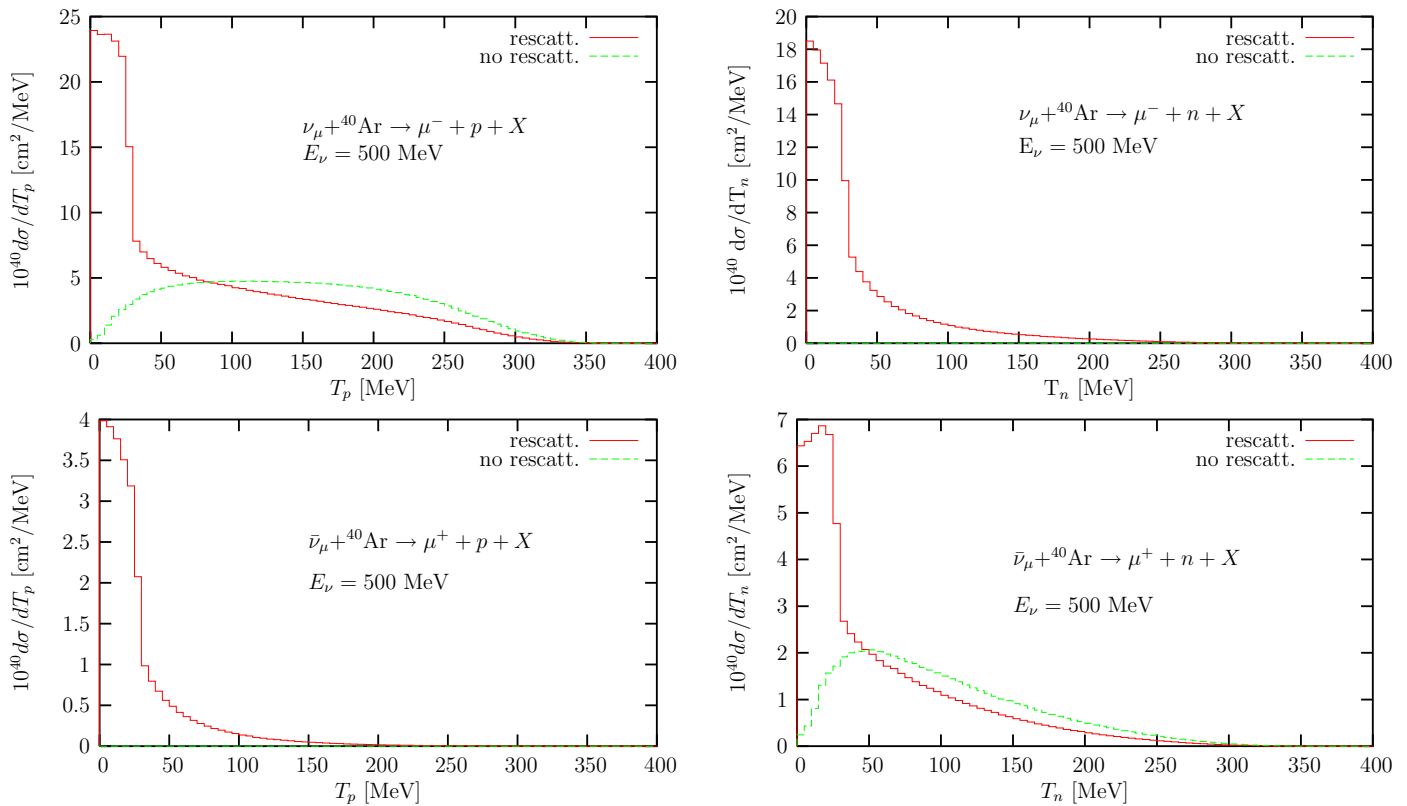}
\caption{  (color online). 
Charged current $^{40}Ar(\nu_{\mu},\mu^-+N)$ (upper panels) and 
$^{40}Ar(\bar{\nu}_{\mu},\mu^++N)$ (lower panels) cross sections as a function
of the kinetic energy of the final nucleon for an incoming neutrino or
antineutrino energy of 500 MeV. Left and right panels correspond to
the emission of protons and neutrons respectively. The dashed histogram 
shows results without nucleon rescattering and the solid one the full model.
}\label{fig:res4}
\end{center}
\end{figure}

The nucleon spectra produced by CC processes induced by muon neutrinos
and antineutrinos of 500 MeV are shown in Fig.~\ref{fig:res4} for
argon. Of course neutrinos only interact via CC with neutrons and
would emit protons, but these primary protons interact strongly with
the medium and collide with other nucleons which
are also ejected. As a consequence there is  a reduction of the flux of high 
energy protons but a large number of secondary nucleons, many of them neutrons,
of lower energies appear. We should recall that our cascade model does not 
include the collisions of nucleons with kinetic energies below 30 MeV. Thus,
the results at those low energies are not meaningful and are shown for 
illustrative purposes only in Fig.\ref{fig:res4}.

The flux reduction due to the quasielastic NN interaction can be easily
accommodated in optical potential calculations. However in those calculations
the nucleons that interact are lost when in the physical process they simply
come off the nucleus with a different energy and angle, and may be charge, and 
they must be taken into account.

\subsubsection{NC Nucleon spectra }

\begin{figure}[tbh]
\begin{center}
\includegraphics[scale=0.7,  bb= 50 350 540 790]{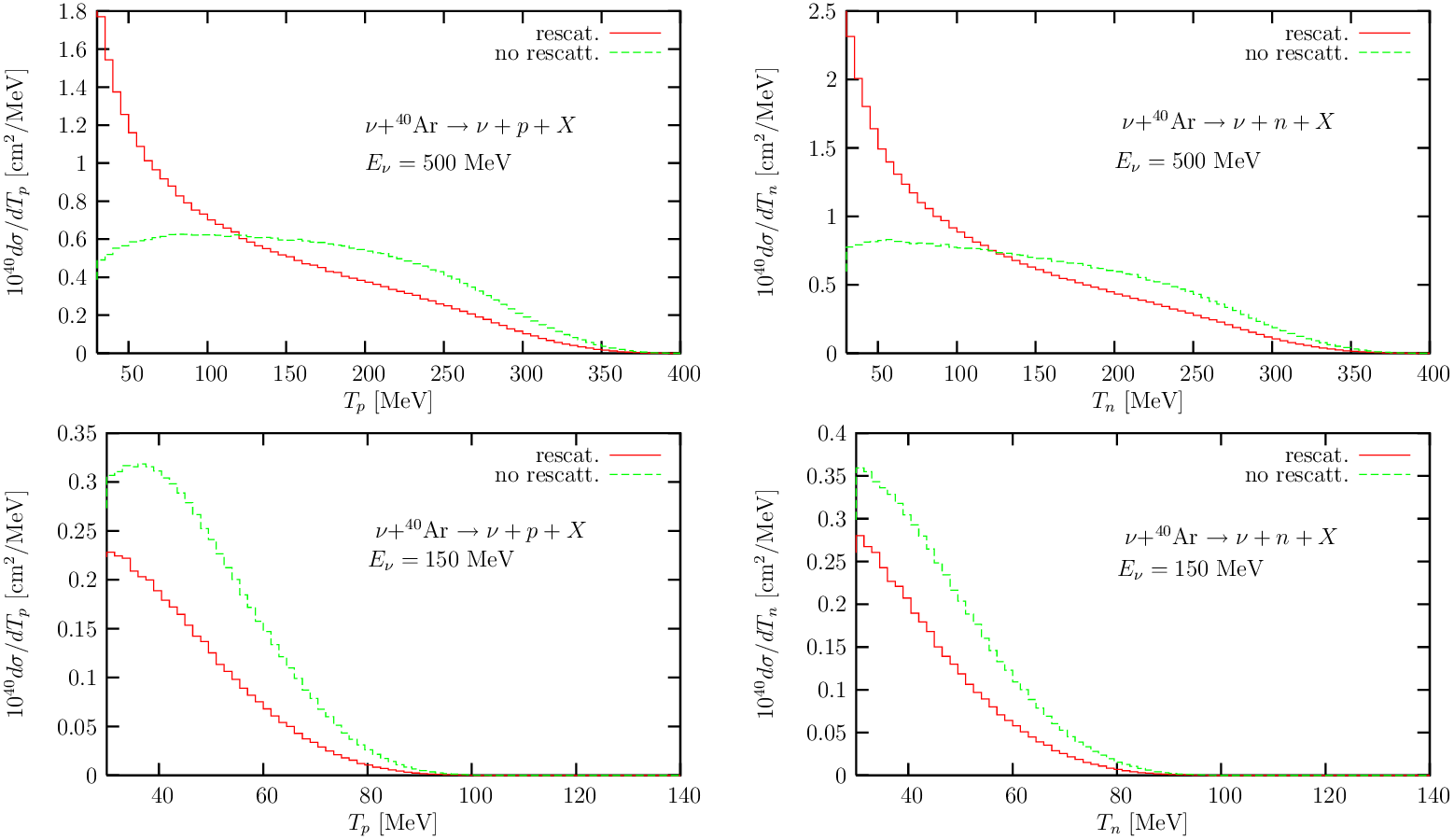}
\caption{ (color online). Neutral current
$^{40}Ar(\nu,\nu+N)$ at 500 MeV (upper panels) and 150 MeV (lower
panels) cross sections as a function of the kinetic energy of the
final nucleon. Left and right panels correspond to the emission of
protons and neutrons respectively. The dashed histogram shows results
without rescattering and the solid one the full model.
}\label{fig:res5}
\end{center}
\end{figure}

\begin{figure}[tbh]
\begin{center}
\includegraphics[scale=0.7,  bb= 50 350 540 790]{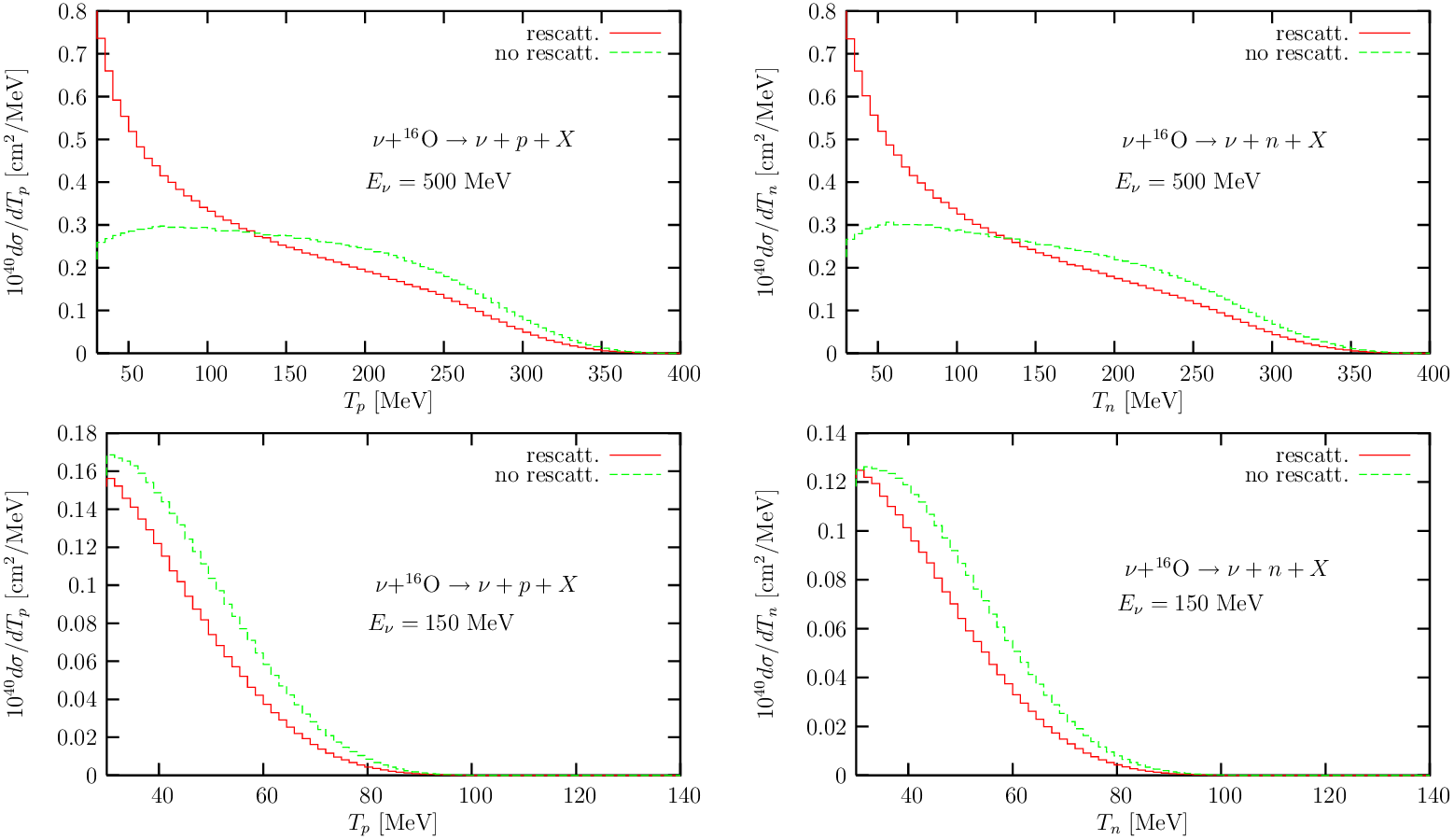}
\caption{ (color online). Same as Fig.\ref{fig:res5} for oxygen.}
\label{fig:res5b}
\end{center}
\end{figure}

The energy distributions of nucleons emitted after a NC interaction are shown
in Figs.~\ref{fig:res5} and \ref{fig:res5b}. In Fig.~\ref{fig:res5}, we show 
the results for $^{40}$Ar  at two different energies. In both cases we find the
large effect of the rescattering of the nucleons.
For 500 MeV neutrinos the rescattering of the outgoing nucleon
produces a depletion of the higher energies side of the spectrum, but
the scattered nucleons clearly enhance the low energies region.  For
lower neutrino energies, most of the nucleons coming from nucleon
nucleon collisions would show up at energies below the 30 MeV cut.

As expected, the rescattering effect is smaller in lighter nuclei as can be seen in
Fig.~\ref{fig:res5b} for oxygen. In all cases the final spectra of 
protons and neutrons are very similar.
Our results without rescattering can be compared with other
calculations like those of
Refs.~\cite{Horowitz:1993rj,vanderVentel:2003km} and other like
Refs.~\cite{Meucci:2004ip,Martinez:2005xe}.  However, in these latter
cases, which incorporate the nucleon final state collisions, via the
use of optical potentials the main effect of rescattering is to reduce
the cross section at all energies instead of displacing the strength
towards lower energies as we find.

Also interesting is the very similar spectrum shape obtained in 
shell model~\cite{vanderVentel:2003km} and Fermi gas calculations, as the 
present one.  

\begin{figure}[tbh]
\begin{center}
\includegraphics[scale=0.8, bb= 50 560 540 790]{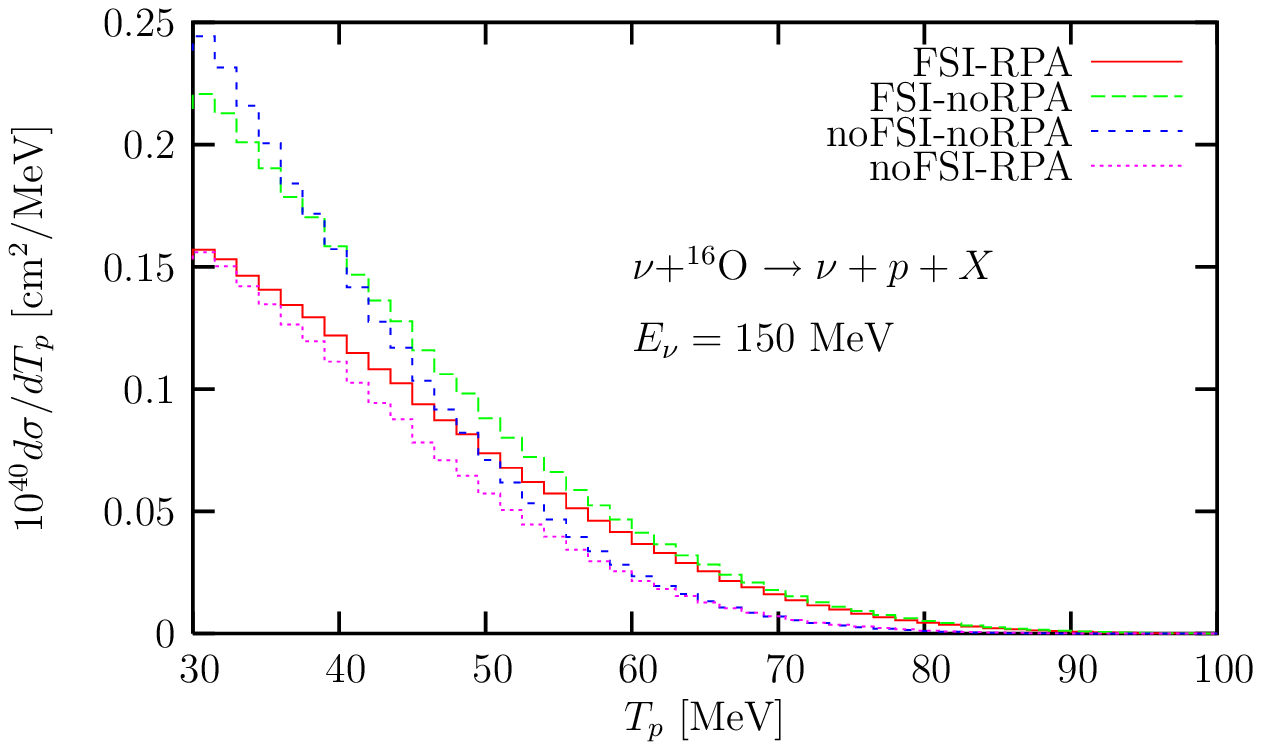}
\caption{ (color online). 
$d\sigma/dT_p'$ cross section for oxygen at
$E_\nu = 150 $ MeV as a function of the kinetic energy of the final
proton for different nuclear models, including or not RPA and FSI
effects. In all cases a MC simulation is performed to compute the
rescattering of the outgoing nucleons. }\label{fig:res6}
\end{center}
\end{figure}
Finally, in Fig.~\ref{fig:res6}, we show the effect over the spectrum
of the RPA correlations and the consideration of the real part of
the nucleon selfenergy (FSI).  We find that the larger reduction due
to RPA takes place for the lower energy nucleons. A similar situation
is obtained for higher energy neutrinos.  The inclusion of FSI effects
produces an enhancement of the number of high energy
protons due to the different nucleon dispersion relation.

\subsubsection{Nucleon angular distributions }
 
\begin{figure}[tbh]
\begin{center}
\includegraphics[scale=0.8, bb= 50 560 540 790]{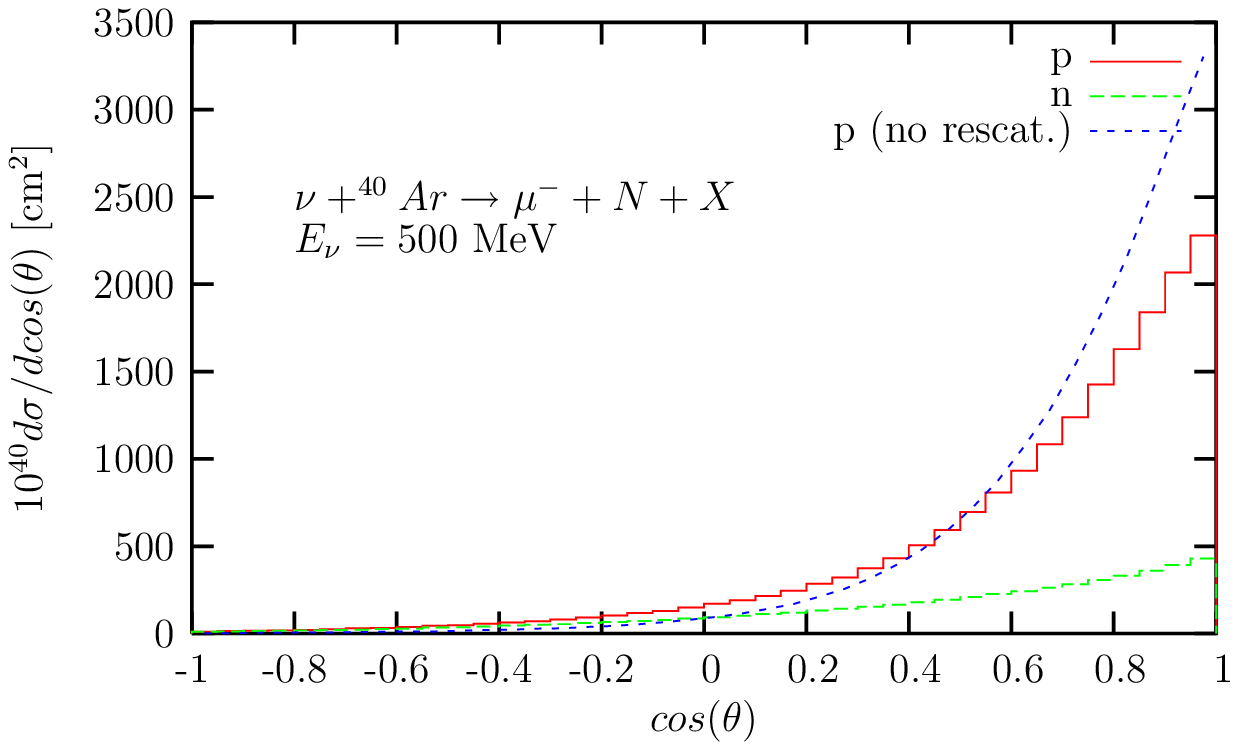}
\caption{ (color online). Nucleon angular distribution for the
 $^{40}Ar(\nu_{\mu},\mu^-+N)$ reaction. The dashed line shows the proton
 results without rescattering and the solid (dotted) histogram stands for 
 proton (neutron) results from our full  model. In all cases, the
 contribution of low energy nucleons (below  30 MeV) is not included.   
 }\label{fig:res7}
\end{center}
\end{figure}

The angular distribution of nucleons is also  affected by the 
rescattering. In Fig.~\ref{fig:res7}, we show  the proton and neutron spectra
in the $\nu_{\mu}+^{40}Ar\to\mu^-+X$ reaction. For comparison we also show the
distribution of protons without rescattering.  
After taking into account the rescattering the proton cross section is less 
forward peaked. Even flatter is the
neutron cross section, because  neutrons come only from secondary NN collisions
and not from the weak neutrino--nucleon interaction.
\begin{figure}[tbh]
\begin{center}
\includegraphics[scale=0.8,  bb= 50 560 540 790]{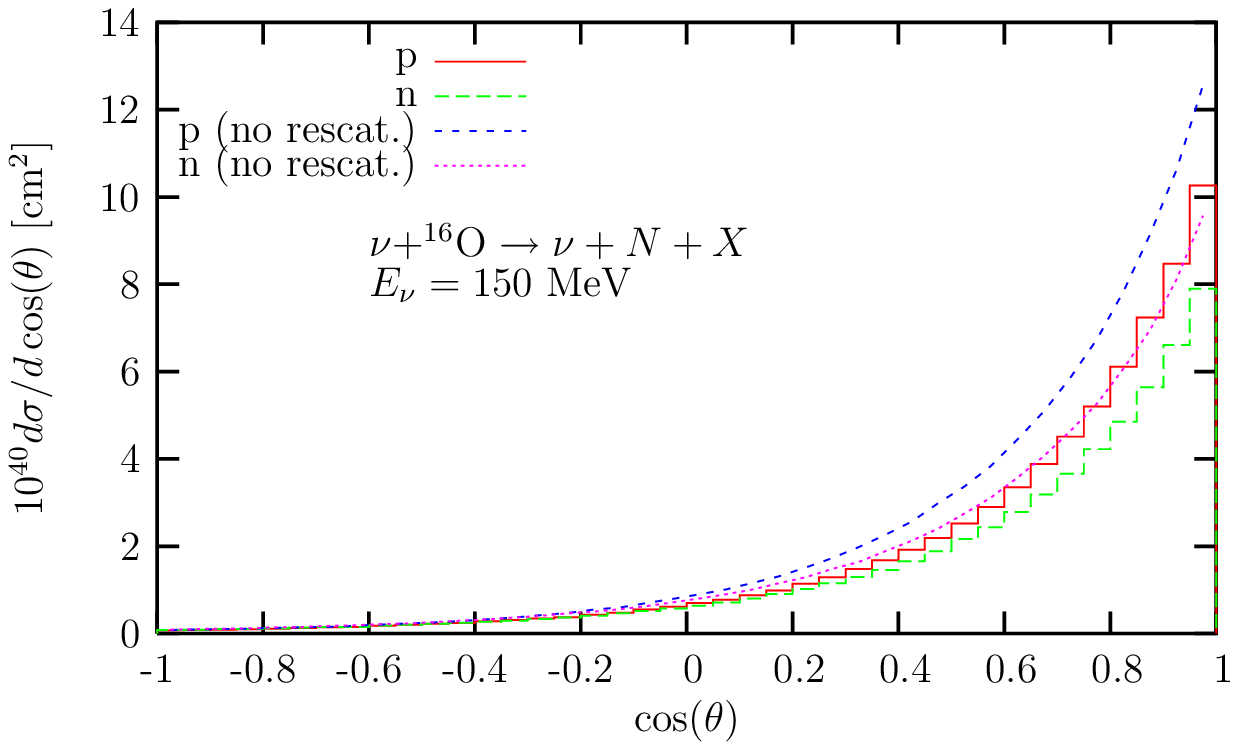}
\caption{ (color online). Proton and neutron angular distribution for the
 $^{16}O(\nu,\nu'+N)$ reaction. In all cases, the contribution of low
 energy nucleons (below 30 MeV) is not included.}\label{fig:res8}
\end{center}
\end{figure}

For NC reactions (Fig.~\ref{fig:res8}), the situation is more symmetric and the
angular distribution is  similar for protons and neutrons, as  was the
case for the energy spectra.

\subsubsection{Energy-Angle distributions}

\begin{figure}[tbh]
\begin{center}\rotatebox{270}{
\includegraphics[scale=0.5]{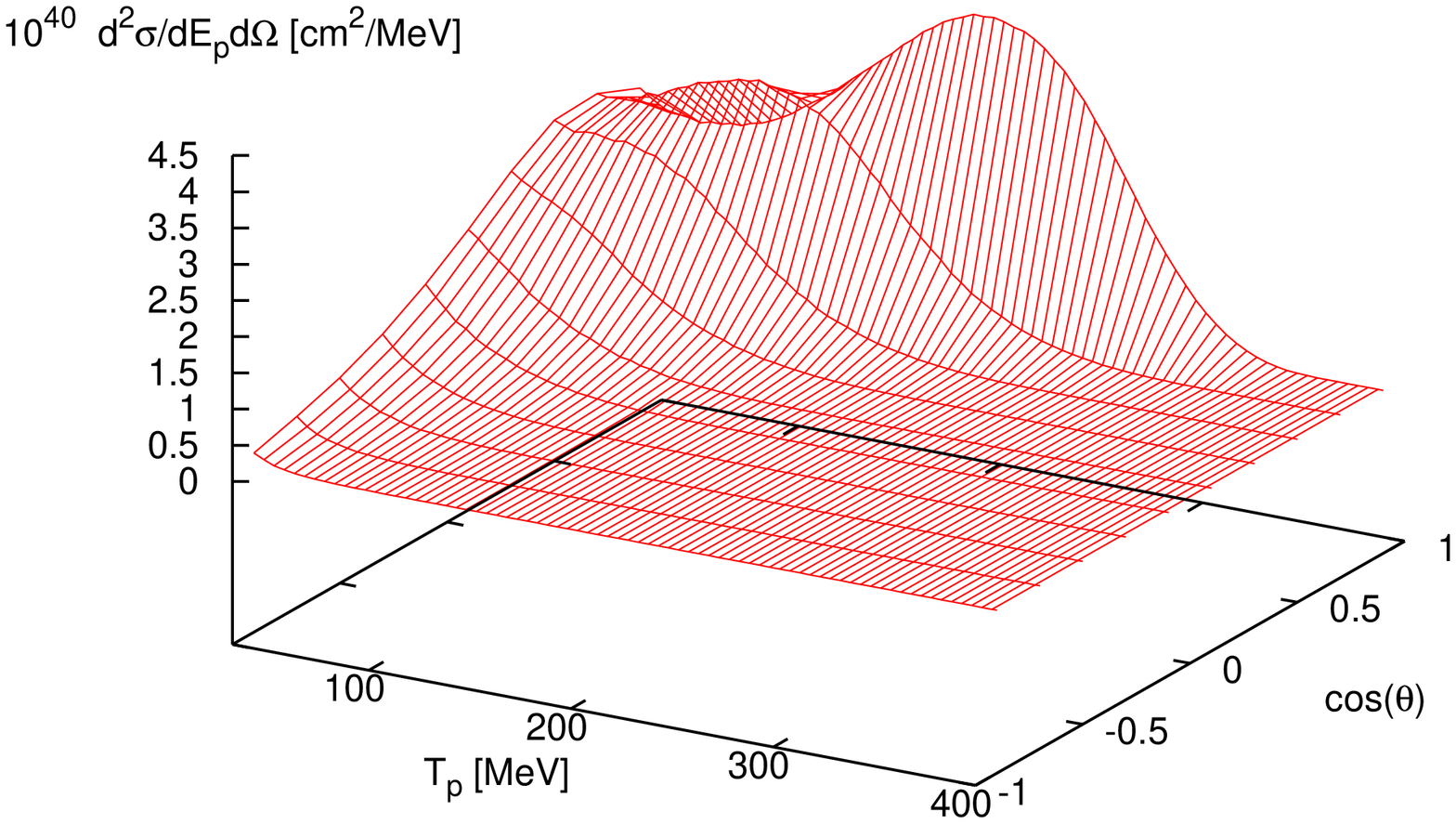}}
\caption{ (color online). Double differential cross section
$d^2\sigma/d\Omega\,dE$ as a function of the outgoing kinetic nucleon
energy and scattering angle for the reaction $\nu_\mu + ^{16}$O$
\to \mu^- + p + X$ and an incoming neutrino energy of 500 MeV.
}\label{fig:res9a}
\end{center}
\end{figure}

\begin{figure}[tbh]
\begin{center}\rotatebox{270}{
\includegraphics[scale=0.5]{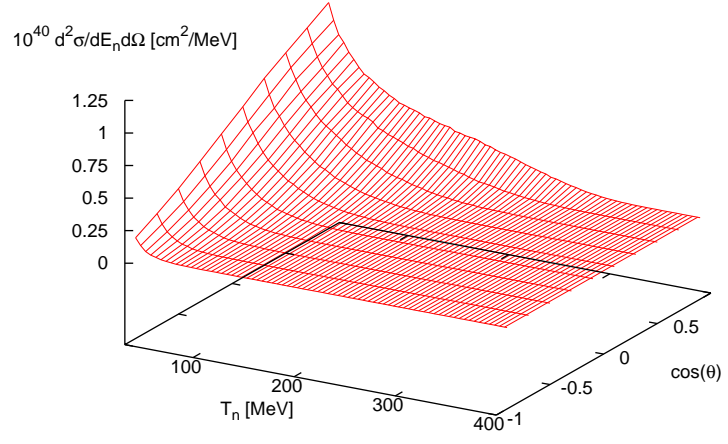}}
\caption{ (color online). Same as Fig.\ref{fig:res9a} for 
$\nu_\mu + ^{16}$O$ \to \mu^- + n + X$ 
}\label{fig:res9b}
\end{center}
\end{figure}

\begin{figure}[tbh]
\begin{center}\rotatebox{270}{
\includegraphics[scale=0.5]{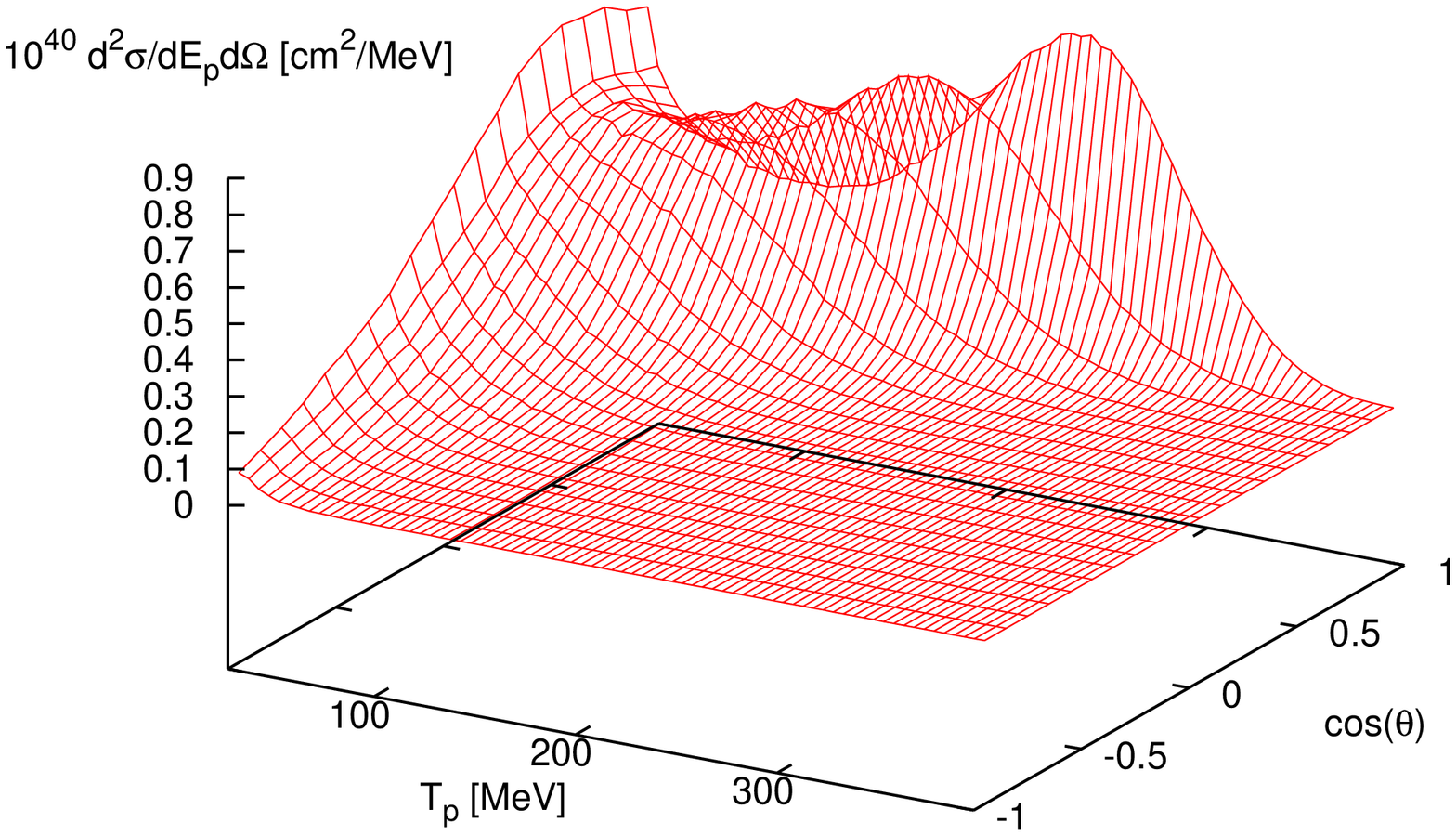}}
\caption{  (color online). Same as Fig.\ref{fig:res9a} for 
$\nu + ^{16}$O$ \to \nu + p + X$. Incoming neutrino energy 500 MeV
}\label{fig:res9c}
\end{center}
\end{figure}

\begin{figure}[tbh]
\begin{center}\rotatebox{270}{
\includegraphics[scale=0.5]{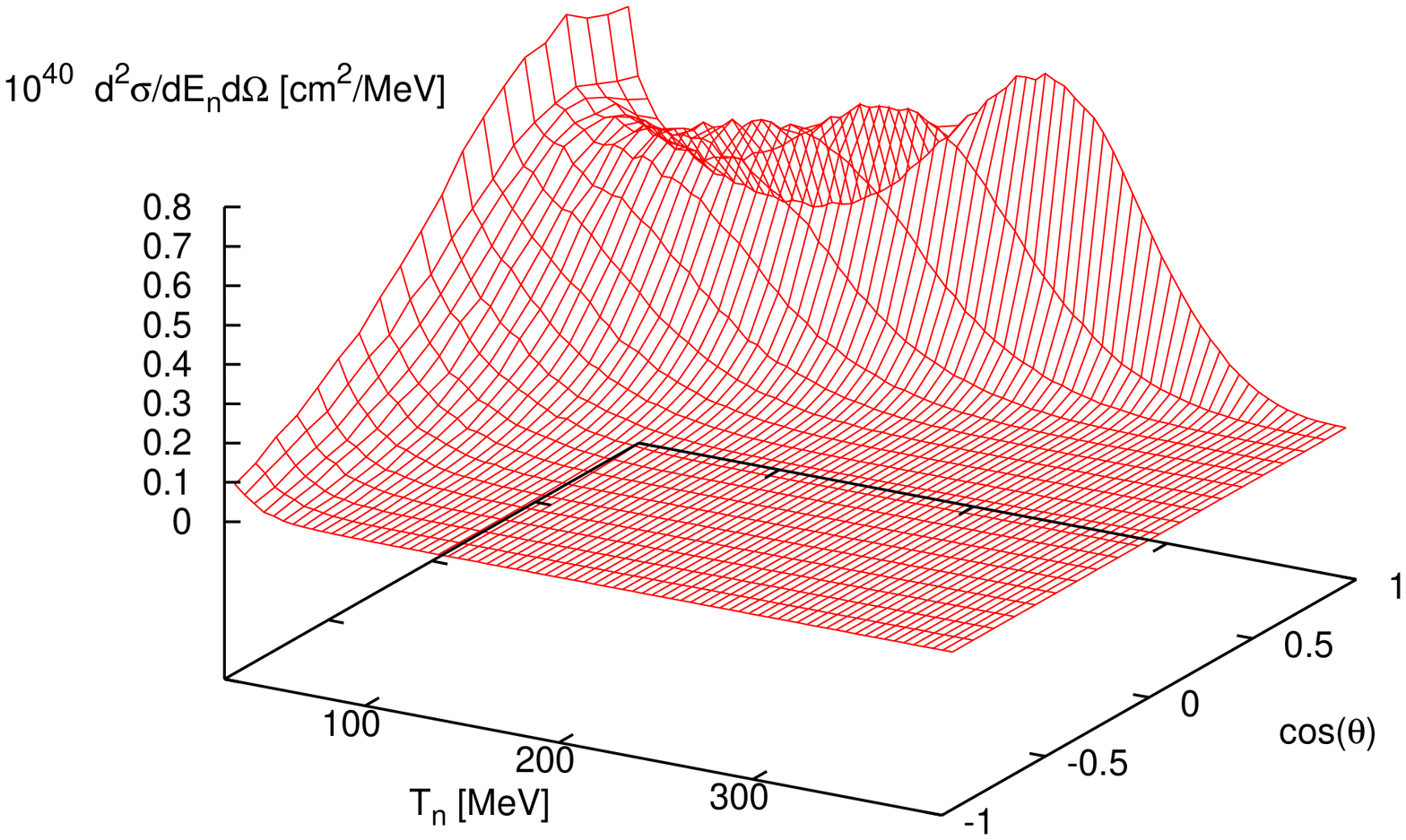}}
\caption{ (color online). Same as Fig.\ref{fig:res9a} for 
$\nu + ^{16}$O$ \to \nu + n + X$. Incoming neutrino energy 500 MeV
}\label{fig:res9d}
\end{center}
\end{figure}

In Fig.~\ref{fig:res9a}--\ref{fig:res9d} we show double differential
energy-angle cross sections for both charged and neutral
currents. Three of the four panels show a common feature with previous
calculations (see i.e. Fig. 1 of Ref.~\cite{Horowitz:1993rj} and
Fig. 5 of Ref.~\cite{vanderVentel:2003km}), namely, at forward angles
there are two peaks, one of them at low energies, that merge into a
single one for larger values of the angle. As it is shown in
Ref.~\cite{vanderVentel:2003km} the use of the nucleon momentum
distributions from shell model wave functions produces less sharp
features at forward angles than the Fermi gas calculation of
Ref.~\cite{Horowitz:1993rj}.
However, our results are even softer than those of
Ref.~\cite{vanderVentel:2003km} although our calculation starts with a
local Fermi gas momentum distribution. The reason is clear, the strong
effects of rescattering change direction and energy of the
nucleons. This can be seen in the panel showing (Fig.~\ref{fig:res9b})
the neutron emission for the CC process
$\nu_\mu+^{16}$O$\to\mu^-+n+X$. In this case all neutrons come from
nucleon-nucleon collisions and spread over the available phase space
without any remarkable feature except the accumulation at low
energies.

\subsubsection{Proton to neutron ratios}

\begin{figure}[tbh]
\centerline{\includegraphics[scale=0.8,  bb= 50 560 540 790]{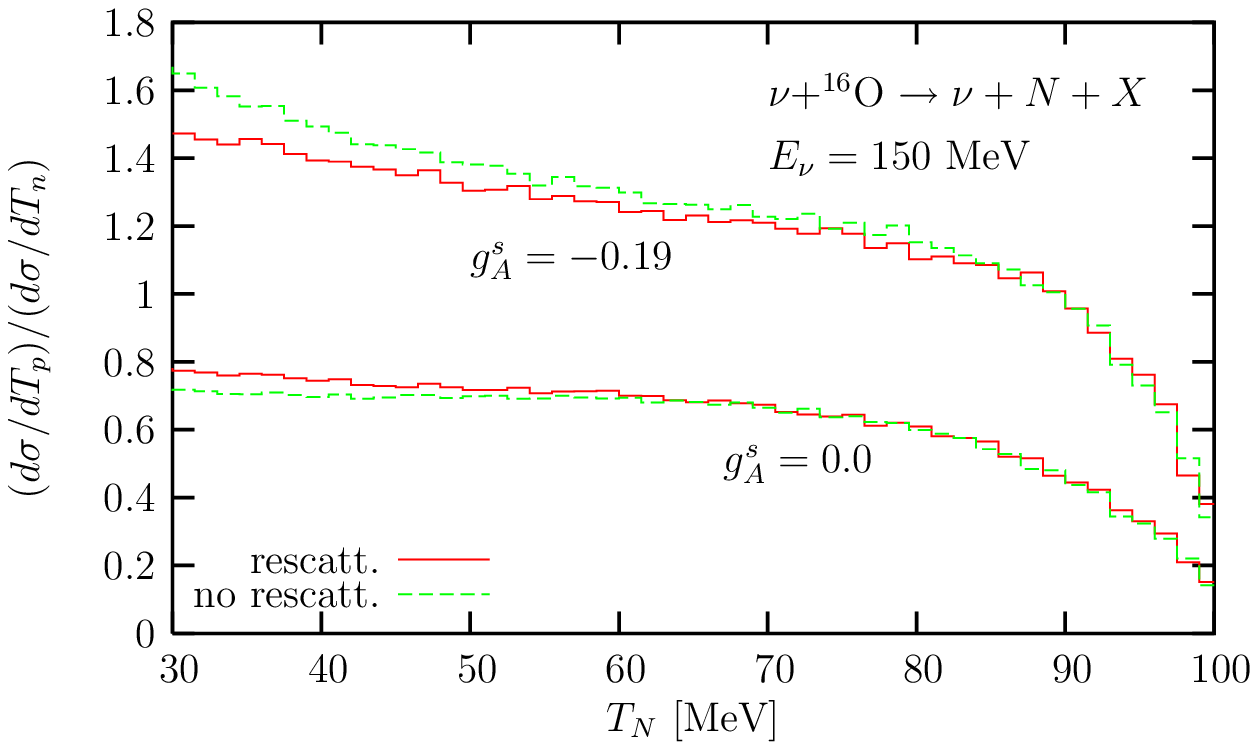}}
\caption{ (color online). Ratio of
$d\sigma/dT$ for protons over that for neutrons for $E_\nu=150$ MeV 
in the reaction $\nu+^{16}O\to \nu'+N+X$
as a function of the nucleon kinetic energy. Dashed histogram: without nucleon rescattering.
Solid histogram: full model.
}\label{fig:res10}
\end{figure}

\begin{figure}[tbh]
\begin{center}
\includegraphics[scale=0.7,  bb= 50 560 540 790]{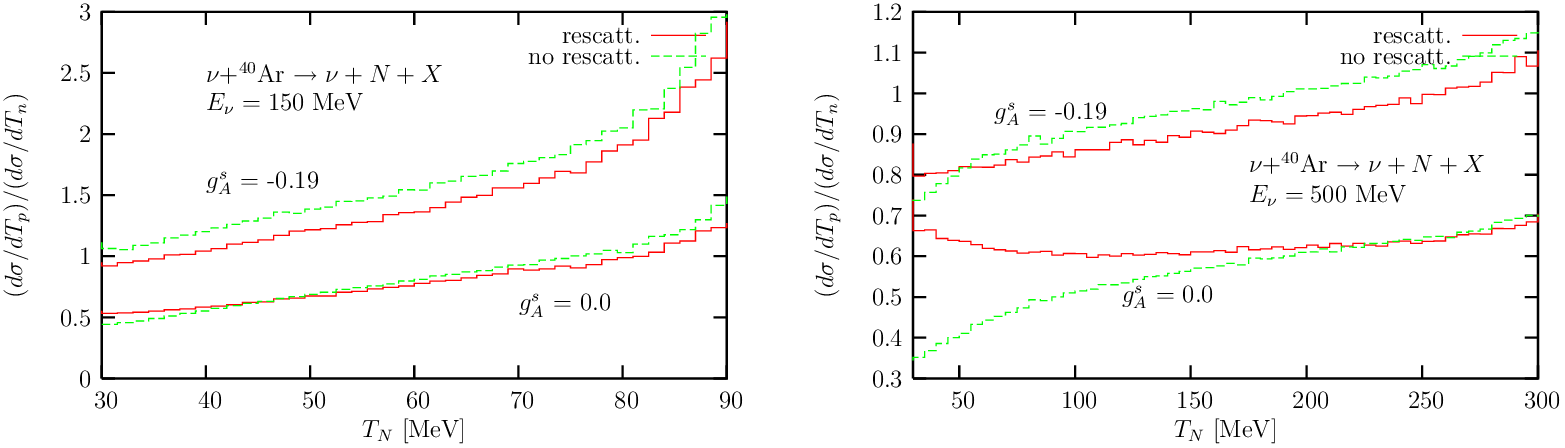}
\caption{ (color online). Ratio of $d\sigma/dE$ for protons over that for
neutrons for $E_\nu=150$ MeV and  $E_\nu=500$ MeV in the reaction
$\nu+^{40}Ar\to \nu'+N+X$ as a function of the nucleon kinetic energy.
Dashed histogram: without nucleon rescattering.
Solid histogram: full model.}
\label{fig:res11}
\end{center}
\end{figure}

The ratio of proton to neutron QE cross section could be very
sensitive to the strange quark axial form factor of the nucleon, and
thus to the $g^s_A$ parameter
\cite{Garvey:1992qp,Garvey:1993sg,Horowitz:1993rj,Alberico98,
vanderVentel:2003km}.  Our results for this ratio in $^{16}$O are
shown in Fig.~\ref{fig:res10}.  We do not consider very low energies
where shell model effects could be more important and our MC
simulation is unreliable.  We find similar results for light nuclei
and low energies as in Ref.~\cite{Garvey:1993sg} where RPA correlations
were taken into account or in Ref.~\cite{Horowitz:1993rj}. Our model
includes as the main additional ingredient the rescattering of the
nucleons via a MC simulation. However also this rescattering
produces minor changes for light nuclei, because of the smaller
average density, and for low energies because most secondary nucleons
are below our 30 MeV cut.
However, the sensitivity to the collisions of the final nucleons 
is larger for both heavier nuclei and for larger
energies of the neutrinos as shown in Fig.~\ref{fig:res11}
where it is clear that one sees the importance of the secondary nucleons at the
low energies side of the spectrum.

\section{Conclusions}
\label{sec:concl}

We have studied the QE contribution to the inclusive $(\nu_l,\nu_l
N)$, $(\nu_l,l^- N)$, $({\bar \nu}_l,{\bar \nu}_l N)$ and $({\bar
\nu}_l,l^+ N)$ reactions in nuclei using a MC simulation
method to account for the collisions of the ejected nucleons during
their way out of the nucleus. As input, we have used the reaction
probability from the microscopical many body framework developed in
Ref.~\cite{NAV05} for CC induced reactions, while for NC we use
results from a natural extension, performed also in this work, of the
model described in that reference. Limitations of the DWIA models have
been discussed. In particular those models cannot properly
describe individual inclusive neutron and proton spectra and for the
ratio of proton $(\nu,p)$ to neutron $(\nu,n)$ yields, the sensitivity
to the collisions of the final nucleons might become important for
both medium and heavy nuclei and for energies of the neutrinos larger
than 150 MeV, as shown in Fig.~\ref{fig:res11}.

\begin{acknowledgments}

J.N. warmly thanks to E.  Oset for various stimulating discussions and
communications.  This work was supported by DGI and FEDER funds,
contracts BFM2002-03218 and BFM2003-00856, by the EU Integrated Infrastructure
Initiative Hadron Physics Project contract RII3-CT-2004-506078
and by the Junta de Andaluc\'\i a.

\end{acknowledgments}


\appendix

\section{NC Nucleon Tensor} 
\subsection{Impulse Approximation}
\label{sec:app_nc}

Taking into
account that in Eq.~(\ref{eq:res}) both the particle and the hole
nucleons are on the mass shell ($p^2=(p+q)^2 = M^2, ~ 2p\cdot q+q^2=0$), one
finds 
\begin{equation}
A_N^{\mu\nu}(p,q) = a_1^N g^{\mu\nu} + a_2^N \left ( p^\mu p^\nu +
\frac{p^\mu q^\nu + p^\nu q^\mu}{2}\right ) + {\rm i} a_3^N
\epsilon^{\mu\nu\alpha\beta}p_\alpha q_\beta + a_4^N q^\mu q^\nu
\end{equation}
with, omitting the obvious subindex $N=n$ or $p$,
\begin{eqnarray}
a_1 (q^2) &=& 8q^2 \left \{ (F_1^Z + \mu_Z F_2^Z)^2 + (G_A^Z)
^2\left (\frac14 -
\frac{M^2}{q^2}\right )  \right \} \nonumber \\
a_2 (q^2) &=& 32 (F_1^Z)^2- 8 (\mu_Z F_2^Z)^2 \frac{q^2}{M^2} + 8
(G_A^Z)^2 \nonumber \\  
a_3(q^2) &=&  16 G_A^Z (F_1^Z+\mu_Z F_2^Z) \nonumber \\ 
a_4(q^2) &=& - \frac{8q^2}{M^2}(\mu_Z F_2^Z)^2 \left (\frac{M^2}{q^2}+
\frac14\right )-16 F_1^Z\mu_Z F_2^Z
 \end{eqnarray}

The cross section for the process $\nu_l +\, N \to \nu_l +
N $  is given by
\begin{equation}
\sigma_{\nu \nu} = \frac{G^2}{32\pi(s-M^2)^2}
\int^{0}_{-(s-M^2)^2/s} dq^2 \Big ( q^2\left\{ a_1  +
    \frac{s}{2}a_2 - \frac{q^2}{2} a_3  \right \} + (s-M^2) \left \{ 
\frac{s-M^2}{2} a_2
    -q^2 a_3 \right \} \Big)
 \label{eq:free}
\end{equation}
where $s=(2|\vec{k}\,|+M)M$ is the Mandelstam variable ($|\vec{k}\,|$
is incoming neutrino energy in the LAB frame). The variable $q^2$ is
related to the outgoing neutrino LAB polar angle ($\theta^\prime$) by
$q^2= (k-k^\prime)^2 = - 2|\vec{k}\,||\vec{k}^\prime\,|(1 -
\cos\theta^\prime)$. 

The cross section for the process $\bar{\nu}_l+N
\to \bar{\nu}_l+N $ is obtained by replacing $a_3$ by $-a_3$.

\subsection {RPA Corrections}
\label{sec:rpaamunu}
Taking $\vec{q}$ in the $z$ direction and after performing the RPA sum
of Fig.~\ref{fig:fig3}, we find, neglecting\footnote{Note that $q^0/M$
is of the order $|\vec{q}\,|^2/M^2$ and as mentioned in
Sect.~\ref{sec:rpa}, we have considered $\mu_Z F_2^Z|\vec{q}\,|/M$ of
order ${\cal O}(0)$.} corrections of order ${\cal
O}\left(k_F\vec{p}^{\,2}/M^2,k_F\vec{p}^{\, \prime 2}/M^2,k_Fq^0/M\right)$
\begin{eqnarray}
\frac{\delta A^{00}_{\rm RPA}}{2M^2} &=& 8 \left
(\frac{E(\vec{p}\,)}{M}\right )^2 \Big \{ ({\bf C_N}-1)
\left[(F_1^Z)^p-(F_1^Z)^n\right]^2 + ({\bf D_N}-1)
\left[(F_1^Z)^p+(F_1^Z)^n\right]^2 \Big \} \nonumber \\ &-&4
\frac{\vec{q}^{\,2}}{M^2} \Big \{ ({\bf C_N}-1)
\left[(F_1^Z)^p-(F_1^Z)^n\right]\left[(\mu_ZF_2^Z)^p-(\mu_ZF_2^Z)^n\right]
\nonumber \\ 
&+& ({\bf D_N}-1) \left[(F_1^Z)^p+(F_1^Z)^n\right]\left[(\mu_ZF_2^Z)^p
+(\mu_ZF_2^Z)^n\right] \Big \} \\
&&\nonumber\\
\frac{\delta A^{0z}_{\rm RPA}}{2M^2} &=&  \left (
\frac{E(\vec{p}\,)}{M}\frac{2p_z+|\vec{q}\,|}{M}\right) \Big \{ 4 
({\bf C_N}-1) \left[(F_1^Z)^p-(F_1^Z)^n\right]^2 + 4 ({\bf D_N}-1)
\left[(F_1^Z)^p+(F_1^Z)^n\right]^2  \nonumber \\ 
&+&  ({\bf C_L}-1) \left[(G_A^Z)^p-(G_A^Z)^n\right]^2 + ({\bf E_N}-1)
\left[(G_A^Z)^p+(G_A^Z)^n\right]^2  \Big\} \\&&\nonumber\\
\frac{\delta A^{zz}_{\rm RPA}}{2M^2} &=& 2({\bf C_L}-1)
 \left[(G_A^Z)^p-(G_A^Z)^n\right]^2 + 2 ({\bf E_N}-1)
\left[(G_A^Z)^p+(G_A^Z)^n\right]^2  \\ &&\nonumber\\
\frac{\delta A^{xx}_{\rm RPA}}{2M^2} &=& -2\frac{q^2}{M^2} \Big \{({\bf
C_T}-1) \left[(\mu_ZF_2^Z)^p-(\mu_ZF_2^Z))^n\right]^2 + ({\bf E_N}-1)
\left[(\mu_ZF_2^Z))^p+(\mu_ZF_2^Z))^n\right]^2 \Big \} \nonumber \\
&-&4\frac{q^2}{M^2} \Big \{ ({\bf C_T}-1)
\left[(F_1^Z)^p-(F_1^Z)^n\right]\left[(\mu_ZF_2^Z)^p-(\mu_ZF_2^Z)^n\right]\nonumber
\\ 
&+& ({\bf E_N}-1)
\left[(F_1^Z)^p+(F_1^Z)^n\right]\left[(\mu_ZF_2^Z)^p+(\mu_ZF_2^Z)^n\right]\Big
\} \nonumber\\
&+& 2({\bf C_T}-1)
 \left[(G_A^Z)^p-(G_A^Z)^n\right]^2 + 2 ({\bf E_N}-1)
\left[(G_A^Z)^p+(G_A^Z)^n\right]^2
 \\ 
&&\nonumber\\
\frac{\delta A^{xy}_{\rm RPA}}{2M^2} &=& -4{\rm i}
\frac{|\vec{q}\,|E(\vec{p}\,)}{M^2} \Big\{ ({\bf
  C_T}-1)\left[(G_A^Z)^p-(G_A^Z)^n\right] 
\left[(F_1^Z+\mu_ZF_2^Z)^p-(F_1^Z+\mu_ZF_2^Z)^n\right] \nonumber \\
&+& ({\bf
  E_N}-1)\left[(G_A^Z)^p+(G_A^Z)^n\right] 
\left[(F_1^Z+\mu_ZF_2^Z)^p+(F_1^Z+\mu_ZF_2^Z)^n\right]  \Big\}
\end{eqnarray}
with the polarization coefficients defined as
\begin{eqnarray}
{\bf C_N}(\rho) &=& \frac{1}{|1-c_0f^\prime_0(\rho)U_N(q,k_F)|^2}, \quad
{\bf C_T}(\rho) = \frac{1}{|1-U(q,k_F)V_t(q)|^2} \quad {\bf
C_L}(\rho) = \frac{1}{|1-U(q,k_F)V_l(q)|^2} \nonumber\\ 
 {\bf D_N}(\rho) &=& 
\frac{1}{|1-c_0f_0(\rho)U_N(q,k_F)|^2}, \quad {\bf E_N}(\rho) =
\frac{1}{|1-c_0g_0(\rho)U_N(q,k_F)|^2}\label{eq:coeffs}
\end{eqnarray}

\end{document}